\documentclass[aps,prb,reprint,amssymb]{revtex4-2}
\usepackage{amsmath,amssymb,color}
\usepackage{bm}
\usepackage{graphicx}
\usepackage{psfrag}
\usepackage{bbold}
\usepackage{bbm}
\usepackage{hyperref}
\newcommand{\bd}{\bm}
\usepackage{pifont}
\begin{document}

\title{Quantum fluctuations in two-dimensional altermagnets \\}

\author{Niklas Cichutek, Peter Kopietz, and Andreas R\"{u}ckriegel}
  
\affiliation{Institut f\"{u}r Theoretische Physik, Universit\"{a}t
  Frankfurt,  Max-von-Laue Stra{\ss}e 1, 60438 Frankfurt, Germany}

\date{\today}

 \begin{abstract}
 
The magnetic properties of two-dimensional altermagnets can be obtained from a square lattice 
Heisenberg model with antiferromagetic nearest neighbor interaction and two types of next-nearest neighbor interactions arranged in a checkerboard pattern. 
Using nonlinear spin-wave theory we calculate for this model the corrections to the renormalized magnon spectrum and the staggered magnetization to first order in the inverse spin quantum number $1/S$. We also show that to order $1/S^2$ the ground state energy is not sensitive to the component of  the interaction which is responsible for altermagnetism.
At the $\Gamma$-point the $1/S$-correction to the magnon dispersion vanishes so that quantum fluctuations do not induce a gap in the magnon spectrum of altermagnets, as expected by Goldstone's theorem. We extract the leading $1/S$-corrections to the spin-wave velocity and the effective mass characterizing the curvature of the magnon dispersion in altermagnets.

\end{abstract}


\maketitle

\section{Introduction}
Altermagnets have gained a lot of attention since their classification in 2022 \cite{Smejkal22a,Smejkal22b,Smejkal23}. These materials exhibit a unique combination of ferromagnetic and antiferromagnetic properties. For example, while they possess zero net magnetization, characteristic of antiferromagnets, they display a spin-polarized electronic band structure typically associated with ferromagnets. These features make them promising candidate materials for spintronics applications. A quantitative understanding of the magnon dispersion is essential for spintronics applications, especially for the development of efficient magnonic devices. Furthermore, the magnon dispersion is relevant for a wide range of phenomena, from thermodynamic quantities to dynamical responses, and it can be probed through both scattering and spectroscopy experiments. 

In a recent work \cite{Cichutek25} we have shown that long-wavelength
magnons in two-dimensional altermagnets can spontaneously decay at zero 
temperature. The decay rate depends on the chirality of the magnons and exhibits an anisotropic momentum dependence reflecting the underlying symmetry.
Spontaneous magnon decay in altermagnets has also been studied in
Refs.~[\onlinecite{Eto25,GarciaGaitan24}].
The calculations in Ref.~[\onlinecite{Cichutek25}] are based on the diagrammatic expansion of the magnon self-energy in powers of the inverse spin quantum number $1/S$ using the effective spin model
for altermagnets proposed by Brekke {\it{et al.}}~\cite{Brekke23}.
Although in Ref.~[\onlinecite{Cichutek25}] we have focused on this specific model, we have argued that the long-wavelength behavior of the decay is generic for a large class of two-dimensional altermagnets provided we substitute the proper values of the spin-wave velocity $c$ and the effective mass $m$ characterizing the curvature of the magnon dispersion.
However, the long-wavelength parameters $c$ and $m$ are renormalized by quantum fluctuations, which we explicitly calculate here  to leading order in $1/S$.
In two dimensions these corrections are finite and do not alter the
qualitative picture of the spontaneous decay at long wavelengths found in 
Ref.~[\onlinecite{Cichutek25}]. 
For realistic values of the model parameters the quantum corrections have the same order of magnitude as the splitting of the magnon energies triggered by altermagnentism.
We also show that the renormalized magnon dispersion remains gapless 
at the $\Gamma$-point in the Brillouin zone, which is of course not surprising because
magnons are the Goldstone modes associated with broken spin-rotational invariance in the antiferromagnetic ground state \cite{Wagner66,Altland10}.

Apart from the renormalized magnon spectrum, we also calculate 
the sublattice magnetization to order $1/S$ and the ground state energy to order $1/S^2$
for the effective altermagnetic spin model \cite{Brekke23}.
Interestingly, to this order the ground state energy is independent of the specific component of the next-nearest neighbor coupling associated with altermagnetism.
For the nearest-neighbor antiferromagnetic Heisenberg model
on a square lattice these corrections have been obtained long time ago by Oguchi \cite{Oguchi60} and our results agree with Ref.~[\onlinecite{Oguchi60}] if we switch off the next-nearest-neighbor couplings in our model.

\section{Checkerboard Heisenberg Model}

\begin{figure}[!htb]
 \begin{center}
 \includegraphics[width=0.45\textwidth]{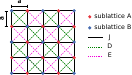}
   \end{center}
  \caption{%
Depiction of the altermagnet spin model given in Eq.~(\ref{eq:modelc}). The spins are located on a two-dimensional equidistant grid with spacing $ a $ and are coupled via a nearest-neighbor antiferromagnetic interaction $J$, leading to the emergence of two sublattices $A$ and $B$. The two next-nearest neighbor interactions $D$ and $E$ connect spins across the diagonals of the plaquettes in a checkerboard pattern. The reduced symmetry due to the  inequivalence of neighboring
plaquettes for
$D \neq E$ is responsible for the specific altermagnetic properties of the model.}
\label{fig:models3}
\end{figure}

To model the magnetic properties of two-dimensional altermagnets we
use the checkerboard Heisenberg model proposed in Ref.~\cite{Brekke23}.
In the notation of Ref.~\cite{Cichutek25} the Hamiltonian is given by
\begin{align}
 {\cal{H}}  = {} & J \sum_{\bd{R} }
  \bd{S}_{\bd{R}} \cdot ( \bd{S}_{ \bd{R} + \bd{a}_1}  +  \bd{S}_{ \bd{R} + \bd{a}_2} )
  \nonumber
  \\
  & 
   + 
   \sum_{\bd{R} \in A} \bd{S}_{\bd{R}} \cdot  (
     D  \bd{S}_{\bd{R} + \bd{d}_1 }
     +
    E  \bd{S}_{\bd{R} + \bd{d}_2 })
   \nonumber
   \\
   &
   +  
   \sum_{\bd{R} \in B}  \bd{S}_{\bd{R}} \cdot (
     E  \bd{S}_{\bd{R} + \bd{d}_1 }
     + D  \bd{S}_{\bd{R} + \bd{d}_2 } ),
  \label{eq:modelc}
 \end{align}
where the spin-$S$ operators $\bd{S}_{\bd{R}}$ are localized at the sites $\bd{R}$ of a two-dimensional square lattice with $N$ sites and lattice spacing $a$.
We divide the lattice into two sublattices A and B
as shown in Fig.~\ref{fig:models3}. 
The model is characterized by three couplings: the antiferromagnetic interaction $J >0$ connects pairs of nearest neighbor spins separated by $\bd{a}_1 = a \hat{\bd{x}}$ or 
$\bd{a}_2 = a \hat{\bd{y}}$, while the next-nearest neighbor interactions
$D$ and $E$ connect spins across the diagonals of the elementary plaquettes separated by $\bd{d}_1 =
a ( \hat{\bd{x}} + \hat{\bd{y}} )$ or $\bd{d}_2 =
a ( - \hat{\bd{x}} + \hat{\bd{y}} )$. 
Here $\hat{\bd{x}}$ and $\hat{\bd{y}}$ are
unit vectors along the crystallographic axes of the lattice.
The couplings $D$ and $E$ are arranged in a checkerboard pattern, as shown in Fig.~\ref{fig:models3}. 
For $D=E$ our model reduces to the so-called $J_1$-$J_2$ model, i.e., 
the square lattice antiferromagnet with nearest-neighbor interaction $J_1 = J$ and next-nearest-neighbor interaction
 \begin{equation}
J_2 = \frac{D + E}{2}.
 \end{equation}
For $D \neq E$ neighboring plaquettes are
inequivalent so that the lattice is covered by two distinct plaquettes; see 
Fig.~\ref{fig:models3}. 
Our model \eqref{eq:modelc} then 
describes a two-dimensional altermagnet. 
To facilitate the comparison with known spin-wave results for the
$J_1$-$J_2$ model \cite{Chandra88} it is convenient to set
 \begin{equation}
 D = J_2 + J_2^\prime, \; \; \; E = J_2 - J_2^\prime,
 \end{equation}
where the energy scale
\begin{equation}
 J_2^\prime = \frac{ D-E}{2}
 \end{equation}
quantifies the strength of the altermagnetism in the system. 
With this notation our Hamiltonian (\ref{eq:modelc}) can also be written as
\begin{align}
 {\cal{H}}  = {} & J \sum_{\bd{R} }
  \bd{S}_{\bd{R}} \cdot ( \bd{S}_{ \bd{R} + \bd{a}_1}  +  \bd{S}_{ \bd{R} + \bd{a}_2} )
  \nonumber
  \\
  & 
   + 
   J_2 \sum_{\bd{R} } \bd{S}_{\bd{R}} \cdot  (
      \bd{S}_{\bd{R} + \bd{d}_1 }
     +
     \bd{S}_{\bd{R} + \bd{d}_2 })
   \nonumber
   \\
   &
   +  J_2^{\prime}
   \sum_{\bd{R} }  e^{ i \bd{Q} \cdot \bd{R} } \bd{S}_{\bd{R}} \cdot (
      \bd{S}_{\bd{R} + \bd{d}_1 }
     -  \bd{S}_{\bd{R} + \bd{d}_2 } ),
  \label{eq:modelc2}
 \end{align}
where $\bd{Q} = ( \pi /a , \pi /a )$ is the antiferromagnetic ordering wavevector so that $e^{ i \bd{Q} \cdot \bd{R} } =1$ for $\bd{R} \in A$ and
$e^{ i \bd{Q} \cdot \bd{R} } =-1$ for $\bd{R} \in B$.
Note that the effective spin model in Eq.~\eqref{eq:modelc2} describes only the magnetic properties
of altermagnets; electronic properties must be modeled by fermionic lattice 
models~\cite{Roig24,Das24,Bose24}.

\section{Spin-wave theory}
\label{sec:spinwavetheory}

\subsection{General procedure}

Now we perform a nonlinear spin-wave expansion around the classical ground state. For sufficiently small altermagnetic couplings relative to the nearest-neighbor interaction $J$, the classical ground state is the N\'eel state. In this case, the red and blue dots in Fig.~\ref{fig:models3} represent classical spin-up and spin-down vectors, respectively, such that two sublattices $A$ and $B$ emerge with spins pointing in opposite directions $\pm \bd{e}_z$. To simplify the notation we enumerate  the lattice sites
${\bd{R}}_i$ by an integer $i = 1, \ldots , N$ and 
write ${\bd{S}}_i = {\bd{S}}_{\bd{R}_i }$.
In the classical N\'eel state the spins 
then point in the directions
\begin{align}
    \bd{m}_i = \zeta_i \bd{e}_z,
\end{align}
where $\zeta_i = e^{ i \bd{Q} \cdot {\bd{R}}_i } = 1 $ for
$ {\bd{R}}_i \in A$ and $\zeta_i = -1$ for
${\bd{R}}_i \in B$.
In order to account for quantum fluctuations, we complement 
$\bd{m}_i$ by two transverse spherical basis vectors
$\bd{e}^{\pm}_i = \bd{e}_x \pm \zeta_i \bd{e}_y$, and represent
the spin operators in this local basis \cite{Schuetz03},
\begin{align}
    \bd{S}_i=S_i^{||} \bd{m}_i + \frac{1}{2} \left( S_i^+ \bd{e}_i^- + S_i^- \bd{e}_i^+ \right),
\end{align}
where $S^{||}_i = \bd{m}_i \cdot {\bd{S}}_i$ and $
\bd{S}_i^{\pm} = \bd{e}_i^{\pm} \cdot \bd{S}_i$.
The spin components are then bosonized using the 
Holstein-Primakoff transformation \cite{Holstein40},
\begin{subequations}
\begin{align}
S_i^+&=\sqrt{2S} \sqrt{1-\frac{a_i^\dagger a_i}{2S}}a_i \approx \sqrt{2S} \left[a_i-\frac{a_i^\dagger a_i a_i}{4S} \right], \\
S_i^-&=\sqrt{2S} a_i^\dagger \sqrt{1-\frac{a_i^\dagger a_i}{2S}} \approx \sqrt{2S} \left[ a_i^\dagger-\frac{a_i^\dagger a_i^\dagger a_i}{4S} \right], \\
S_i^{||}&=S-a_i^\dagger a_i.
\end{align}
\end{subequations}
The expansion of the square roots  up to the first order 
is sufficient to obtain the $1/S$-correction to the magnon dispersion and the sublattice magnetization, and the $1/S^2$-correction of the ground state energy.
The bosonized Hamiltonian is then expanded in powers of $1/S$,
\begin{align}
\label{eq:hphamiltonian}
    \mathcal{H}=\mathcal{H}_0+\mathcal{H}_2+\mathcal{H}_4+ \ldots,
\end{align}
where $\mathcal{H}_{2n}$ is of order $S^{2-n}$. To zeroth order we obtain the classical ground state energy
\begin{align}
\mathcal{H}_0 = & -2NJS^2 + N(D+E)S^2 = - 2 N ( J - J_2) S^2.
\label{eq:Hcl}
\end{align}
To write down the quadratic part ${\cal{H}}_2$ of the bosonized Hamiltonian
we transform to momentum space,
\begin{align}
    a_{i}= \begin{cases}
  \sqrt{\frac{2}{N}} \sum\limits_{\bd{k}} e^{i \bd{k} \cdot \bd{R}_i} a_{\bd{k}}, \qquad \bd{R}_i \in A,      \\   \sqrt{\frac{2}{N}} \sum\limits_{\bd{k}} e^{i \bd{k} \cdot \bd{R}_i} b_{\bd{k}}, \qquad \bd{R}_i \in B,
\end{cases}
\end{align}
where the momentum sum is over the reduced Brillouin zone associated with the antiferromagnetic ground state. We obtain
\begin{align}
 &\mathcal{H}_2{}={}  4 JS \sum\limits_{\bd{k}} \Bigg\{ \left[1-\frac{J_2}{J}(1-\lambda_{\bd{k}}^A) \right] a^\dagger_{\bd{k}} a_{\bd{k}} \nonumber\\
& \, + \left[ 1-\frac{J_2}{J}(1-\lambda_{\bd{k}}^B)\right] b^\dagger_{\bd{k}} b_{\bd{k}}  + \gamma_{\bd{k}} (a^\dagger_{\bd{k}} b^\dagger_{-\bd{k}} + b_{-\bd{k}} a_{\bd{k}})\Bigg\},
\label{eq:H2}
\end{align}
where
\begin{subequations}
\begin{align}
\lambda_{\bd{k}}^A &= \frac{D}{D+E} \cos (\bd{k} \cdot \bd{d}_1) + \frac{E}{D+E} \cos (\bd{k} \cdot \bd{d}_2), \\
\lambda_{\bd{k}}^B &= \frac{E}{D+E} \cos (\bd{k} \cdot \bd{d}_1) + \frac{D}{D+E} \cos (\bd{k} \cdot \bd{d}_2), \\
\gamma_{\bd{k}} &=\frac{1}{2} \left[ \cos (k_x a) + \cos (k_y a) \right].
\end{align}
\end{subequations}
Finally, taking Umklapp scattering into account, the quartic part of the bosonized Hamiltonian can be written as
\begin{widetext}
   \begin{align}
\mathcal{H}_4 & {}={} \frac{2}{N} \sum\limits_{\bd{1}\bd{2}\bd{3}\bd{4}\bd{G}} \delta_{\bd{1}+\bd{2}+\bd{3}+\bd{4},\bd{G}} \bigg\{ W_{\bd{2}+\bd{3}}\Big[a_{-\bd{1}}^\dagger b_{-\bd{2}}^\dagger b_{\bd{3}} a_{\bd{4}} + s_{\bd{G}} b^\dagger_{-\bd{1}}a^\dagger_{-\bd{2}}a_{\bd{3}} b_{\bd{4}} \Big]
+\frac{1}{2!} W_{\bd{4}} \Big[ a_{-\bd{1}}^\dagger a_{\bd{2}} a_{\bd{3}} b_{\bd{4}} + b_{-\bd{4}}^\dagger a_{-\bd{3}}^\dagger a_{-\bd{2}}^\dagger a_{\bd{1}} \nonumber\\
&\phantom{+\frac{1}{2!} W_{\bd{4}} \Big[}+ s_{\bd{G}} \left[ a^\dagger_{-\bd{4}}b^\dagger_{-\bd{3}}b^\dagger_{-\bd{2}} b_{\bd{1}} + b^\dagger_{-\bd{1}} b_{\bd{2}} b_{\bd{3}} a_{\bd{4}} \right] + \frac{1}{(2!)^2} \left[ W^A_{\bd{1} \bd{2} ; \bd{3} \bd{4}} a_{-\bd{1}}^\dagger a^\dagger_{-\bd{2}} a_{\bd{3}} a_{\bd{4}} + W^B_{\bd{1} \bd{2} ; \bd{3} \bd{4}} s_{\bd{G}}
b_{-\bd{1}}^\dagger b_{-\bd{2}}^\dagger b_{\bd{3}}^\dagger a_{\bd{4}} \right] \bigg\},
\label{eq:h4holstein}
\end{align} 
\end{widetext}
where
\begin{subequations}
\begin{align}
    W_{\bd{k}} {}={} &  - 2 J \gamma_{\bd{k}},\\
    W^A_{\bd{1}\bd{2};\bd{3}\bd{4}} {}={}& 2 J_2 \Big[\lambda^A_{\bd{1}+\bd{3}}+\lambda^A_{\bd{1}+\bd{4}}+\lambda^A_{\bd{2}+\bd{3}}+\lambda^A_{\bd{2}+\bd{4}} \nonumber\\
    &\hspace{0.2cm} -\lambda^A_{\bd{1}}-\lambda^A_{\bd{2}}-\lambda^A_{\bd{3}}-\lambda^A_{\bd{4}} \Big],\\
    W^B_{\bd{1}\bd{2};\bd{3}\bd{4}} {}={}& 2 J_2 \Big[\lambda^B_{\bd{1}+\bd{3}}+\lambda^B_{\bd{1}+\bd{4}}+\lambda^B_{\bd{2}+\bd{3}}+\lambda^B_{\bd{2}+\bd{4}} \nonumber\\
    & \hspace{0.2cm} -\lambda^B_{\bd{1}}-\lambda^B_{\bd{2}}-\lambda^B_{\bd{3}}-\lambda^B_{\bd{4}} \Big],
\end{align}
\end{subequations}
To avoid double subscripts we have abbreviated the momentum labels $\bd{k}_1, \bd{k}_2 , \ldots$ by $\bd{1}, \bd{2} ,
\ldots $, 
and $ \bd{G} $ are the reciprocal lattice vectors of sublattice $ A $,
defined by
\begin{equation}
e^{ i \bd{G} \cdot \bd{R} } = 1 
\text{ for } 
\bd{R} \in A .
\end{equation}
The sum over
$\bd{G}$ in Eq.~\eqref{eq:h4holstein} takes Umklapp processes into account, which decorate some of the vertices
with additional minus signs contained in the factor
 $s_{\bd{G}}=\text{sign}( \gamma_{\bd{G}})$. In general, the phase factors associated with
 Umklapp scattering cannot be neglected 
in spin-wave calculations of non-universal quantities which are determined
by short-wavelength fluctuations \cite{Smit20}. 
To completely diagonalize the quadratic part \eqref{eq:H2} of our bosonized Hamiltonian
 we introduce new boson operators $\alpha_{\bd{k}}$ and $\beta_{\bd{k}}$ via the canonical (Bogoliubov) transformation
\begin{align}
    \begin{pmatrix}
        a_{\bd{k}} \\
        b_{-\bd{k}}^\dagger
    \end{pmatrix}
    =\begin{pmatrix}
        u_{\bd{k}} & -v_{\bd{k}} \\
        -v_{\bd{k}}^* & u_{\bd{k}}^*
    \end{pmatrix}
    \begin{pmatrix}
        \alpha_{\bd{k}} \\
        \beta_{-\bd{k}}^\dagger
    \end{pmatrix}.
    \label{eq:Bogoliubov}
\end{align}
The coefficients $u_{\bd{k}}$ and $v_{\bd{k}}$ are chosen such that the new operators $\alpha_{\bd{k}}$ and $\beta_{\bd{k}}$ satisfy the canonical commutation relations, which implies
 $|u_{\bd{k}}|^2-|v_{\bd{k}}|^2=1$.
Furthermore, the requirement that the quadratic part of the transformed Hamiltonian
does not contain any anomalous terms implies
\begin{subequations}
\begin{align}
    u_{\bd{k}}&=e^{i\varphi_{\bd{k}}} \sqrt{\frac{1-\eta_{\bd{k}}+\epsilon_{\bd{k}}}{2\epsilon_{\bd{k}}}},\\
    v_{\bd{k}}&=e^{-i\varphi_{\bd{k}}} \text{sign}(\gamma_{\bd{k}})\sqrt{\frac{1-\eta_{\bd{k}}-\epsilon_{\bd{k}}}{2\epsilon_{\bd{k}}}},
\end{align}
\label{eq:uv}
\end{subequations}
where $\varphi_{\bd{k}}$ is an arbitrary phase and
\begin{subequations} \label{eq:eta_epsilon}
\begin{align}
    \eta_{\bd{k}}&=\frac{J_2}{J}[1-    \cos (k_x a) \cos (k_y a)   ], \\
    \epsilon_{\bd{k}}&=\sqrt{(1-\eta_{\bd{k}})^2-\gamma_{\bd{k}}^2}.
\end{align}
\end{subequations}
In terms of the Bogoliubov bosons $\alpha_{\bd{k}}$ and $\beta_{\bd{k}}$
the
quadratic part ${\cal{H}}_2$ of the bosonized Hamiltonian
defined in Eq.~\eqref{eq:H2}  can then be written as
\begin{equation}
\mathcal{H}_2 = E_0^{(1)} + \sum\limits_{\bd{k}} \left[ \omega_{\bd{k}}^{+} \alpha^\dagger_{\bd{k}} \alpha_{\bd{k}} + \omega_{\bd{k}}^{-} \beta^\dagger_{\bd{k}} \beta_{\bd{k}} \right],
\label{eq:H2diag}
\end{equation}
where the bare energy dispersion of magnons with chirality $p = \pm$ is given by \cite{Cichutek25, Brekke23}
\begin{align}
 {\omega_{\bd{k}}^{p}}={}& 4 J S \epsilon_{\bd{k}} - 4 p J_2^\prime S \sin ( k_x a ) \sin ( k_y a ),
\end{align}
and the constant $E_0^{(1)}$ can be
identified with the leading quantum correction to the classical 
ground state energy, see Eq.~\eqref{eq:e1} below.
Although the  Hamiltonian $ {\cal{H}}_2 + {\cal{H}}_4$
defined in Eqs.~\eqref{eq:H2} and \eqref{eq:h4holstein}
is normal-ordered in terms of the Holstein-Primakoff bosons $a_{\bd{k}}$ and
$b_{\bd{k}}$, this is not the case any more if we  express the Hamiltonian 
in terms of the Bogoliubov bosons $\alpha_{\bd{k}}$ and $\beta_{\bd{k}}$.
The explicit expression for ${\cal{H}}_4$ in terms of Bogoliubov bosons is given
in Appendix~A.
Using the canonical commutation relations of the Bogoliubov bosons to restore normal ordering generates extra terms that encode quantum fluctuations.
In particular, normal ordering of $\mathcal{H}_2$ generates the leading quantum correction
$E_0^{(1)}$ 
of relative order $1/S$ to the ground state energy, while normal ordering $\mathcal{H}_4$
generates the next-to-leading $1/S^2$-correction to the ground state energy and the
leading $1/S$-correction to the magnon dispersion.

\subsection{Staggered Magnetization}

\begin{figure}[tb]
 \begin{center}
  \centering
 \includegraphics[width=0.45\textwidth]{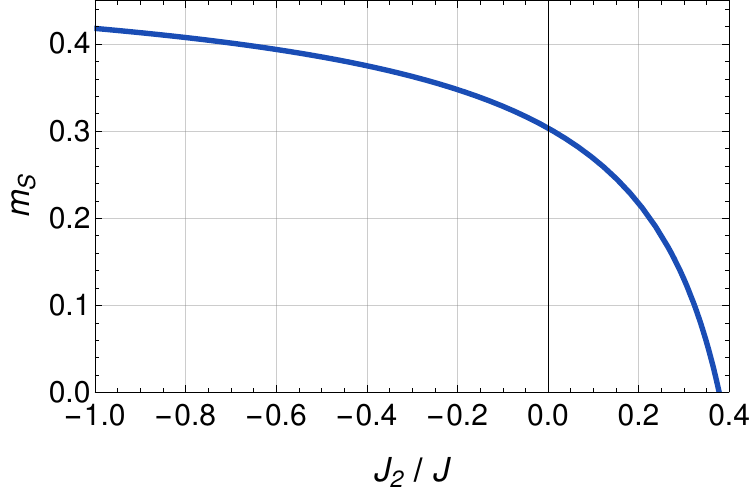}
   \end{center}
  \caption{%
Staggered magnetization $m_s$ as a function of $J_2/J$ obtained from the numerical evaluation of the sum in Eq.~\eqref{eq:staggeredmagnetization} for $S=1/2$. 
Here and below we evaluate the lattice sums in the thermodynamic limit $N \rightarrow 
\infty$ where the sums can be replaced by  integrals over the reduced magnetic Brillouin zone. 
The magnetization decreases with increasing coupling ratio and is zero at $J_2/J\approx 0.378$,
signaling the definite breakdown of spin-wave theory \cite{Chandra88}.}
\label{fig:staggeredmagnetization}
\end{figure}

Before addressing the corrections to the ground state energy and the magnon dispersion,
let us calculate the staggered (or sublattice) magnetization to estimate the regime where our truncated $1/S$-expansion is accurate.
Spin-wave theory is expected to be accurate even for small $S$ 
provided the corrections to the classical approximation for the
staggered magnetization $m_s \approx S$ are small. 
Denoting by $| E_0 \rangle$ the exact ground state of the system,
the staggered magnetization at zero temperature is given by
\begin{align}
    m_s & =  \frac{1}{N} \sum\limits_{i } \langle E_0 | S_i^{||} | E_0 \rangle
 =    
     S - \frac{1}{N} \sum_{i} \langle E_0 |a_i^\dagger a_i | E_0 \rangle .
\end{align}
After Fourier transformation and Bogoliubov transformation, this can be written as
\begin{align}
    m_s 
&=S- \frac{2}{N} \sum_{{\bd{k}}} \langle E_0 | \left( |u_{\bd{k}}|^2 \alpha_{\bd{k}}^\dagger \alpha_{\bd{k}} +|v_{\bd{k}}|^2 \beta_{-\bd{k}} \beta_{-\bd{k}}^\dagger \right. \nonumber \\
& \left. \phantom{=S- aa} - u^*_{\bd{k}} v_{\bd{k}} \alpha_{\bd{k}}^\dagger \beta_{-\bd{k}}^\dagger - u_{\bd{k}} v^*_{\bd{k}} \beta_{-\bd{k}} \alpha_{\bd{k}}  \right) | E_0 \rangle  .
\end{align}
To leading order in $1/S$ the ground state $| E_0 \rangle $ can be approximated by the vacuum of the Bogoliubov bosons defined by
$\alpha_{\bd{k}} |E_0 \rangle=0 = 
\beta_{\bd{k}} |E_0 \rangle$, so that
\begin{align}
    m_s = S - \frac{2}{N} \sum_{{\bd{k}}} |v_{\bd{k}}|^2 + {\cal{O}} \left( \frac{1}{ S} \right).
    \label{eq:staggeredmagnetization}
\end{align}
Note that the Bogoliubov coefficients defined via Eqs.~\eqref{eq:uv} and \eqref{eq:eta_epsilon},
and hence the staggered magnetization to this order, 
depend only on the
ratio $J_2 / J$ and are independent of the altermagnetic energy scale $J_2^\prime = ( D - E )/2$. 
A numerical evaluation of the truncated $1/S$-expansion
\eqref{eq:staggeredmagnetization} of the staggered magnetization as a function of $J_2 /J$ 
is shown in Fig.~\ref{fig:staggeredmagnetization}.
In the limit $J_2/J \to - \infty$ the staggered magnetization approaches its classical value of $S$ and decreases monotonically as $J_2 /J$ increases. For $S=1/2$ the leading $1/S$-correction cancels the classical value at $J_2/J \approx 0.378$ so that 
spin-wave theory definitely breaks down \cite{Chandra88}. In fact, the $1/S$-expansion loses its validity a bit earlier when the quantum correction significantly reduces the magnitude of the classical magnetic moment.
For a square lattice antiferromagnet with only nearest neighbor coupling ($J_2 = J_2^\prime =0$)
the coefficient of the correction of order $1/S$ to $m_s$ vanishes \cite{Castilla91}, so that
the leading correction to Eq.~\eqref{eq:staggeredmagnetization} is actually  of order
$1/S^2$. On the other hand, for finite next-nearest neighbor coupling  the coefficient of order $1/S$
in the expansion \eqref{eq:staggeredmagnetization} is expected to be finite; the calculation of this correction is beyond the scope of this work.

\subsection{Ground state energy}

\begin{figure}[tb]
 \begin{center}
  \centering
\hspace{5mm} \includegraphics[width=0.41\textwidth]{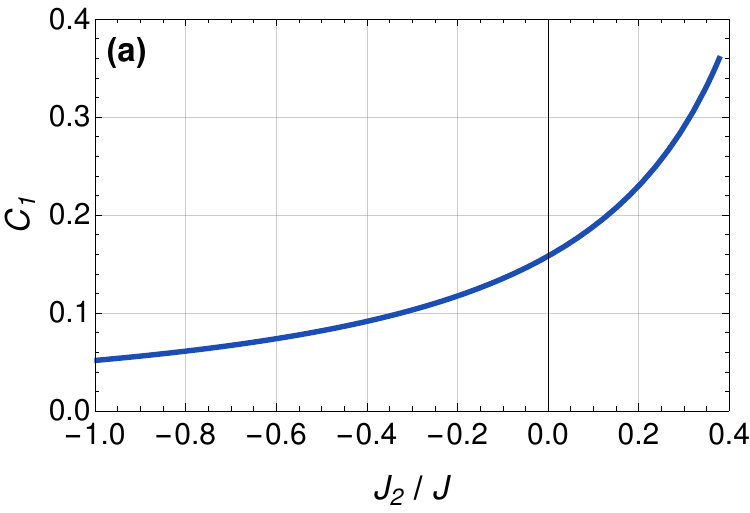}\\ \vspace{5mm}
 \includegraphics[width=0.45\textwidth]{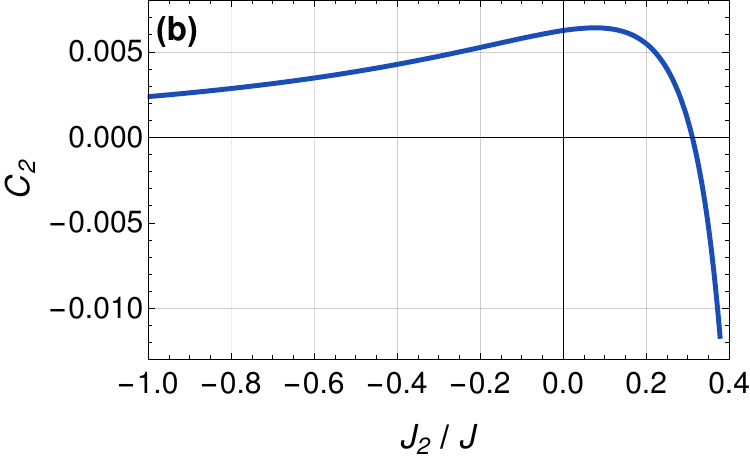}
   \end{center}
  \caption{%
Coefficients of (a) $1/S$- and (b) $1/S^2$-corrections to the ground state energy defined by Eq.~\eqref{eq:ECC}
as a function of $J_2/J$. 
Note that the $1/S^2$-correction is two orders of magnitude smaller than the leading $1/S$-corrections,
which is itself an order of magnitude smaller than the classical ground state energy.}
\label{fig:energycorrection}
\end{figure}

Let us now discuss the leading $1/S$-correction and the next-to-leading
$1/S^2$-correction to the
ground state energy $E_0$ of our model. These corrections can be obtained by normal ordering the
bosonized Hamiltonian after Bogoliubov transformation, retaining terms up to fourth order in the bosons.
Technical details are given in Appendix~B.
We obtain
\begin{align}
    E_0 = E_0^{(0)} + E_0^{(1)} + E_0^{(2)} +
 {\cal{O}} \left( \frac{1}{S} \right),    
    \label{eq:energycorr}
\end{align}
where the classical ground state energy  per lattice site is
according to Eq.~\eqref{eq:Hcl} given by
 \begin{equation} 
 \frac{E_0^{(0)}}{N}  = -2 ( J-J_2 ) S^2,
  \end{equation}
while for the leading correction of order $S$ we find 
\begin{align}
    \frac{E_0^{(1)}}{N}  &= -2 J S \frac{2}{N} \sum\limits_{\bd{k}} (1-\eta_{\bd{k}} -\epsilon_{\bd{k}}).
    \label{eq:e1}
\end{align}
The next correction to the ground state energy is of order $S^0 = 1$ and can be obtained by expressing the quartic part ${\cal{H}}_4$ of the bosonized Hamiltonian in terms of Bogoliubov bosons $\alpha_{\bd{k}}$ and $\beta_{\bd{k}}$ and then normal ordering.
We obtain
    \begin{align}
    \frac{E_0^{(2)}}{N}  = & -\frac{J}{2}  I_1^2 -\frac{J_2}{2}  I_2^2 ,
    \label{eq:energycorrection2}
\end{align}
where
the dimensionless integrals $I_1$ and $I_2$ are given by
\begin{subequations} \label{eq:I12def}
 \begin{align}
 I_1 & =  1- \frac{2}{N} \sum\limits_{\bd{k}} \frac{1-\eta_{\bd{k}}-\gamma_{\bd{k}}^2}{\epsilon_{\bd{k}} } ,
 \label{eq:I1def}
 \\
 I_2 & = \frac{2}{N}  \sum\limits_{\bd{k}} \frac{1-\eta_{\bd{k}}-\epsilon_{\bd{k}}}{\epsilon_{\bd{k}}} (1-\lambda_{\bd{k}}).
 \label{eq:I2def}
 \end{align}
 \end{subequations}
Recall that $\epsilon_{\bd{k}}$ and $\eta_{\bd{k}}$ depend on $J_2 / J$,
so that the right-hand side of Eq.~\eqref{eq:energycorrection2} is a nonlinear function of $J_2$. However, to this order in $1/S$ the ground state energy does not depend
on the altermagnetic energy scale $J_2^\prime = ( D-E)/2$.
For $J_2=0$ the corrections \eqref{eq:e1} and \eqref{eq:energycorrection2} reduce to the 
known expressions for the nearest-neighbor antiferromagnet on a square lattice~\cite{Oguchi60}.
To illustrate the relative magnitude of the corrections as functions of $J_2 /J$, we plot in 
Fig.~\ref{fig:energycorrection} the dimensionless coefficients $C_1$ and $C_2$ defined 
by
\begin{equation}
\frac{E_0}{E_0^{(0)}}  = 1 + \frac{C_1}{S} + \frac{C_2}{S^2}  + {\cal{O}}
\left( \frac{1}{S^3} \right).
\label{eq:ECC}
\end{equation}
Because in the regime where spin-wave theory is valid
$E_0^{(0)}$ is negative while $C_1$ and $C_2$ are both positive, we conclude that quantum fluctuations of spin waves lower the ground state energy, independently of the altermagnetic energy scale $J_2^\prime = ( D-E )/2$. 
Note also that the $ 1 / S $-corrections to the ground state energy rapidly decrease with increasing order,
indicating the reliability of the $ 1 / S $-expansion in this regime.

\subsection{Renormalized magnon energies}

We now calculate the renormalized magnon energies to first order in $1/S$. In contrast to the corrections to the sublattice magnetization and the ground state energy discussed above, 
the quantum corrections to the renormalized magnon energies depend also on the
altermagnetic energy scale $J_2^\prime =  (D - E )/2$. 
If we express ${\cal{H}}_2 + {\cal{H}}_4$ in terms of Bogoliubov bosons and normal order the resulting expressions, we obtain
\begin{equation}
{\cal{H}}_2 + {\cal{H}}_4 = E_0^{(1)} + E_0^{(2)} +  : {\cal{H}}^\prime_2  : + 
  : {\cal{H}}_4 :,
 \end{equation}
where $: \ldots :$ denotes normal-ordering with respect to the Bogoliubov bosons
$\alpha_{\bd{k}}$ and $\beta_{\bd{k}}$. 
The normal-ordered interaction
$: {\cal{H}}_4 :$ is explicitly given in Eq.~\eqref{eq:quarticHamiltonian} of 
Appendix~A,
and the normal-ordered quadratic Hamiltonian $: {\cal{H}}_2^{\prime} :$ differs from
the quadratic part of $ {\cal{H}}_2 $  given in Eq.~\eqref{eq:H2diag} by additional self-energy corrections arising from the normal-ordering of the interaction ${\cal{H}}_4$, 
\begin{align}
: \mathcal{H}^\prime_2 :  =  \sum\limits_{\bd{k}} \Big[ &
( \omega_{\bd{k}}^{+} + \Sigma^+_{\bd{k}} ) \alpha^\dagger_{\bd{k}} \alpha_{\bd{k}} + ( \omega_{\bd{k}}^{-} + \Sigma^-_{\bd{k}} ) \beta^\dagger_{\bd{k}} \beta_{\bd{k}}\nonumber\\  
&+\Delta_{\bd{k}} \left( \alpha_{- \bd{k}} \beta_{\bd{k}} +  \beta^\dagger_{\bd{k}} 
 \alpha^\dagger_{- \bd{k}} \right)\Big].
\label{eq:quadrHamiltonian}
\end{align}
After some lengthy algebra outlined in Appendix~B we find for the diagonal self-energies 
 \begin{align}
  &  {\Sigma_{\bd{k}}^{p}} = 2 I_1 J \frac{1-\eta_{\bm{k}} - \gamma_{\bm{k}}^2}{\epsilon_{\bm{k}}}  \nonumber\\
    & + 2 I_2 \left[ J_2  \frac{1-\eta_{\bm{k}}}{\epsilon_{\bm{k}}} \left( 1-\lambda_{\bm{k}}\right) + p J_2^\prime \sin (k_x a) \sin (k_y a)  \right],
    \label{eq:dispcorr}
\end{align}
and for the off-diagonal self-energy
\begin{align}
    \Delta_{\bd{k}}= & -   \frac{ 2 \gamma_{\bd{k}}}{\epsilon_{\bd{k}} }
    \left[   I_1 J  \eta_{\bd{k}}   
     +  I_2 J_2   
    (1-\lambda_{\bd{k}})  \right],
    \label{eq:xi}
\end{align}
where $I_1$ and $I_2$ are defined in Eqs.~\eqref{eq:I12def}.
For $J_2 = J_2^\prime =0$ the off-diagonal self-energy $\Delta_{\bd{k}}$ vanishes while
the diagonal self-energies $\Sigma_{\bd{k}}^p$ are independent of the chirality label $p$ and reduce to the well-known result for the nearest-neighbor antiferromagnet \cite{Oguchi60}.

\begin{figure}[tb]
\centering
\includegraphics[width=0.45\textwidth]{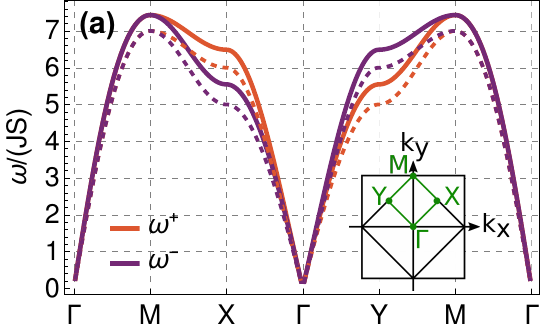}\\
\vspace{4mm}
\includegraphics[width=0.45\textwidth]{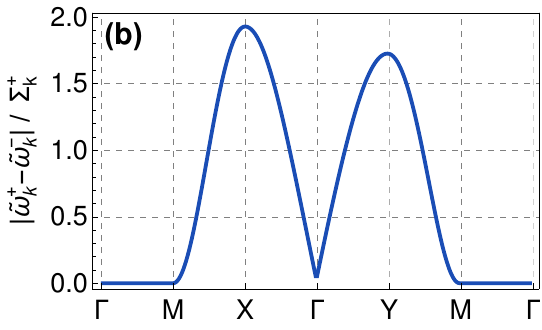}
  \vspace{4mm}
  \caption{
(a) Renormalized magnon dispersions ${\tilde{\omega}_{\bd{k}}^{p}}$ defined
in  Eq.~\eqref{eq:omegaren} for $S=1/2$,
 $J_2/J = - 3/8$, $J_2^\prime/J = -1/8$.  The Brillouin zone path is illustrated in the inset, where the high-symmetry points are labeled by capital letters, for example
  $M=(0,\pi /a)$. Dashed lines represent the leading-order dispersions $\omega_{\bd{k}}^p$ and solid lines represent the renormalized dispersions $\tilde{\omega}_{\bd{k}}^p$ including the leading $1/S$-corrections. (b) Ratio of the energy splitting
$ | \tilde{\omega}_{ \bm{k} }^+ 
  - \tilde{\omega}_{ \bm{k} }^- | $ 
of the renormalized magnon dispersions \eqref{eq:omegaren} relative to the $S^0$-correction 
$ \Sigma_{ \bm{k} }^+ $,
given in Eq.~\eqref{eq:dispcorr}, 
to the magnon dispersion with chirality $p=+$. 
The plot is for the same path in the Brillouin zone and the same parameters as in (a).}
\label{fig:dispersionscorr}
\end{figure}

\begin{figure}[tb]
 \begin{center}
  \centering
 \includegraphics[width=0.45\textwidth]{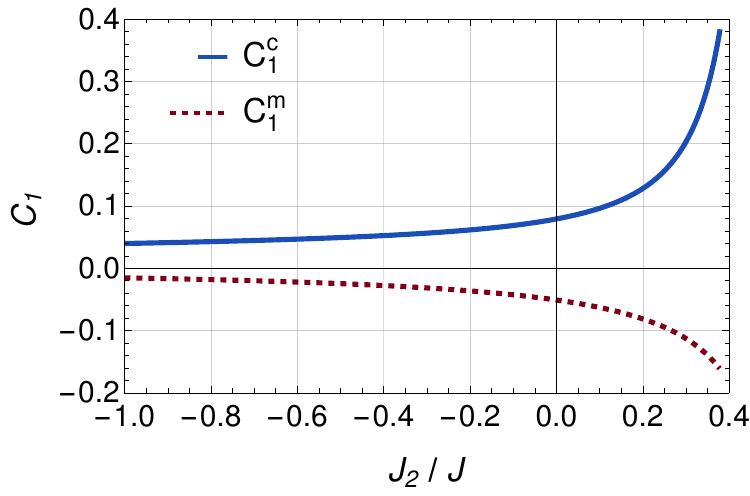}
   \end{center}
  \caption{%
Coefficients $C_1^c$ and $C_1^m$ of the leading $1/S$-corrections 
to the magnon velocity $c$ and the inverse magnon mass $1/m$ as a function of $J_2/J$;
see Eqs.~\eqref{eq:ccorrection} and \eqref{eq:cmcorre}.}
\label{fig:ccorr}
\end{figure}

To obtain the renormalized magnon spectrum, we should now diagonalize the modified quadratic Hamiltonian using another
Bogoliubov transformation. 
The renormalized magnon dispersions are given by 
 \begin{align}
 \tilde{\omega}_{\bd{k}}^p = {} & \frac{p}{2} \left( \omega_{\bd{k}}^+ 
 - \omega_{\bd{k}}^- + \Sigma_{\bd{k}}^+  - \Sigma_{\bd{k}}^- \right)
 \nonumber
 \\
 & + \sqrt{ \frac{1}{4} \left( \omega_{\bd{k}}^+   + \omega_{\bd{k}}^-  + \Sigma_{\bd{k}}^+
   + \Sigma_{\bd{k}}^- \right)^2 - \Delta^2_{\bd{k}} }.
  \end{align}
Expanding this in powers of $1/S$ and keeping 
in mind that $\omega_{\bd{k}}^p = {\cal{O}} ( S )$ while
$\Sigma_{\bd{k}}^p$ and $\Delta_{\bd{k}}$ are of order $S^0 = 1$, 
we conclude that
 \begin{align}
 \tilde{\omega}_{\bd{k}}^p = \omega_{\bd{k}}^p + \Sigma^p_{\bd{k}} +
  {\cal{O}} \left( \frac{1}{S} \right),
  \end{align}
so that the off-diagonal self-energy $\Delta_{\bd{k}}$ does not contribute
to the leading correction to the magnon dispersion.
Hence,  the renormalized
magnon dispersion of our checkerboard Heisenberg model is
 \begin{align}
 \tilde{\omega}^p_{\bd{k}} = {} &
  4 J S \epsilon_{\bd{k}} - 4 p J_2^\prime S \sin ( k_x a ) \sin ( k_y a )
   \nonumber
   \\
   & + 2 I_1 J \frac{1-\eta_{\bm{k}} - \gamma_{\bm{k}}^2}{\epsilon_{\bm{k}}} 
   + 2 I_2  J_2  \frac{1-\eta_{\bm{k}}}{\epsilon_{\bm{k}}} \left( 1-\lambda_{\bm{k}}\right) 
    \nonumber\\
    & + 2 p  I_2  J_2^\prime \sin (k_x a) \sin (k_y a) 
  + {\cal{O}} \left( \frac{1}{S} \right).    
 \label{eq:omegaren}
 \end{align}
Likewise,
normal ordering of 
$ \mathcal{H}_2 + \mathcal{H}_4 $
after the second Bogoliubov transformation only generates corrections of order
$ 1 / S $ 
to the ground state energy and magnon dispersions.
In Fig.~\ref{fig:dispersionscorr}~(a), we show the magnon dispersions with and without $S^0$-correction for the indicated path in the Brillouin zone and $S=1/2$. 
The ratio of the energy splitting
$ | \tilde{\omega}_{ \bm{k} }^+ 
  - \tilde{\omega}_{ \bm{k} }^- | $  
of the renormalized magnon dispersions relative to the $S^0$-correction  $\Sigma_{\bd{k}}^+ $ to the renormalized magnon dispersion with chirality $p=+$  
is displayed in Fig.~\ref{fig:dispersionscorr}~(b),
for the same path in the Brillouin zone and the same set of parameters.
Note that for the choice of the parameters specified  in the caption 
of Fig.~\ref{fig:dispersionscorr} the renormalization  due to quantum fluctuations is larger or has the same order of magnitude as the splitting of the magnon bands induced by the altermagnetic coupling $J_2^\prime$ throughout the whole Brillouin zone. Thus, 
it is important for a quantitative description of the magnon dispersions. 
On the other hand, 
the magnitude of the splitting is barely changed by the $S^0$-correction to the dispersion. 
Note also that the leading $1/S$-correction is momentum dependent and therefore cannot be taken into account by rescaling the leading-order result by a momentum-independent factor. 
Of particular interest is the long-wavelength limit where the magnon
dispersions take on the following universal form \cite{Cichutek25,Lundemo2025}:
 \begin{align}
 \omega^{\pm}_{\bd{k}} & = c | \bd{k} | \pm \frac{ k_x k_y}{m} + {\cal{O}} ( k^3 ),
 \label{eq:omegaexp}
 \end{align}
where $c$ is the magnon velocity and $1/m$ is an inverse mass. The fact that the dispersion remains gapless at the $\Gamma$-point, even after including the $1/S$-corrections, serves as a nontrivial consistency check in 
light of Goldstone's theorem \cite{Wagner66,Altland10}.
Expanding the renormalized dispersion \eqref{eq:omegaren} in powers of momentum, 
we obtain for the  magnon velocity
 \begin{align}
     c=2 \sqrt{2} J S a \sqrt{ 1 - \frac{2J_2}{J} } \left( 1+\frac{C_1^c}{S} \right),
\end{align}
where
\begin{align}
    C_1^c = &  \frac{   I_1 +  ( I_2 - I_1 ) \frac{J_2 }{ J } }{2 \left( 1 - 2 \frac{J_2}{  J}  \right) } .
    \label{eq:ccorrection}
\end{align}
Similarly, we obtain for the renormalized effective mass
    \begin{align}
        \frac{1}{m}= -4 J_2^\prime S a^2 \left( 1+\frac{C_1^m}{S} \right),
    \end{align}
where
    \begin{align}
        C_1^m=  - \frac{I_2}{2} .
        \label{eq:cmcorre}
    \end{align}
A graph of the coefficients $C_1^c$ and $C_1^m$  as functions of $J_2 / J$ is shown in 
Fig.~\ref{fig:ccorr}.
Both coefficients vanish in the limit  $J_2/J \to -\infty$ where quantum fluctuations are completely suppressed.  
Note furthermore that quantum fluctuations increase the magnon velocity,
but decrease the inverse mass.
In Ref.~\cite{Cichutek25}, it was shown that the magnons in this model can undergo spontaneous decay. The linear spin-wave dispersion, characterized by the long-wavelength parameters $c$ and $1/m$, was used to demonstrate that the decay depends on chirality, and an explicit expression for the decay rate in terms of $c$ and $1/m$ was derived. Since the $1/S$-corrections presented in Eq.~\eqref{eq:omegaren} merely renormalize the parameters $c$ and $1/m$, the results of the $1/S$-expansion can be directly incorporated into the analysis of Ref.~\cite{Cichutek25}.\\\\

\section{Summary and conclusions}

In this work we have
used   nonlinear spin-wave theory to study quantum fluctuations in a minimal effective spin  model
(checkerboard Heisenberg model with alternating next-nearest neighbor couplings
$D$ and $E$) for two-dimensional altermagnets \cite{Brekke23}.
We have calculated the leading $1/S$-corrections to the sublattice magnetization and the 
magnon dispersions. While the correction to the sublattice magnetization depends 
only on the average $J_2 = ( D +  E )/2$ of the next-nearest neighbor couplings, the momentum dependence of the
leading quantum correction to the magnon dispersions is rather complicated and
depends also on the energy scale $J_2^\prime = ( D - E ) /2$ associated with altermagnetism. In the long-wavelength limit quantum corrections do not affect the gapless nature of the magnons, in agreement with Goldstone's theorem. We have explicitly calculated the leading quantum corrections to the magnon velocity $c$ and the
inverse magnon mass $1/m$ which characterize the model-indepenent long-wavelength dispersion of altermagnetic magnons. In the regime where quantum fluctuations are not too strong so that the staggered magnetization is not significantly suppressed, the quantum corrections to $c$ and $1/m$ are rather small, even for $S=1/2$. 

We have also calculated the first two quantum corrections (up to relative order $1/S^2$) to the ground state energy of our model. Surprisingly, up to this order the ground state energy is still independent of the altermagnetic coupling $J_2^\prime$. We expect that higher orders in the $1/S$-expansion will eventually generate some dependence of the
ground state energy on $J_2^\prime$, 
which will give a hint whether altermagnetism 
can be stabilized by quantum fluctuations.

\section*{ACKNOWLEDGEMENTS}
We thank the Deutsche Forschungsgemeinschaft (DFG, German Research Foundation) for financial support via TRR 288 - 422213477.\\\\

\begin{appendix}
\setcounter{equation}{0}
\renewcommand{\theequation}{A\arabic{equation}}
\section*{APPENDIX A:  
Quartic Part of the bosonized Hamiltonian}

After Bogoliubov transformation and normal ordering
with respect to  the Bogoliubov bosons  $\alpha_{\bd{k}}$ and $\beta_{\bd{k}}$
the normal-ordered quartic part of
our bosonized Hamiltonian can be written as
\begin{widetext}
\begin{align}
: \mathcal{H}_4 : {}& ={}\frac{2}{N} \sum\limits_{\bd{1}\bd{2}\bd{3}\bd{4}} \sum\limits_{\bd{G}} \delta_{\bd{1}+\bd{2}+\bd{3}+\bd{4},\bd{G}} \bigg\{ \frac{1}{(2!)^2} \Gamma_{\bd{G}}^{(1A)}(\bd{3},\bd{4};\bd{2},\bd{1}) \alpha^\dagger_{-\bd{3}} \alpha^\dagger_{-\bd{4}} \alpha_{\bd{2}} \alpha_{\bd{1}} +
\frac{1}{(2!)^2} \Gamma_{\bd{G}}^{(1B)}(\bd{3},\bd{4};\bd{2},\bd{1}) 
: \beta_{\bd{3}} \beta_{\bd{4}}   \beta^\dagger_{-\bd{2}} \beta^\dagger_{-\bd{1}}  : \nonumber\\
&+\frac{1}{2!} \Gamma_{\bd{G}}^{(2A)}(\bd{3};\bd{4};\bd{2},\bd{1}) \alpha^\dagger_{-\bd{3}} \beta_{-\bd{4}} \alpha_{\bd{2}} \alpha_{\bd{1}}
+\frac{1}{2!} \Gamma_{\bd{G}}^{(2B)}(\bd{3};\bd{4};\bd{2},\bd{1}) 
: \beta_{\bd{3}} \alpha^\dagger_{-\bd{4}}
\beta^\dagger_{-\bd{2}} \beta^\dagger_{-\bd{1}}: +\frac{1}{2!} \Gamma_{\bd{G}}^{(3A)}(\bd{3},\bd{4},\bd{2};\bd{1}) \alpha^\dagger_{-\bd{3}} \alpha^\dagger_{-\bd{4}} \alpha_{\bd{2}} \beta^\dagger_{-\bd{1}}\nonumber\\
&+\frac{1}{2!} \Gamma_{\bd{G}}^{(3B)}(\bd{3},\bd{4},\bd{2};\bd{1}) 
: \beta_{\bd{3}} \beta_{\bd{4}}
\beta^\dagger_{-\bd{2}}  \alpha_{\bd{1}} : 
+\frac{1}{2!} \Gamma_{\bd{G}}^{(4A)}(\bd{3};\bd{4};\bd{2};\bd{1}) 
: \alpha^\dagger_{-\bd{3}}  \beta_{\bd{4}} \beta^\dagger_{-\bd{2}}  \alpha_{\bd{1}} : +\frac{1}{2!} \Gamma_{\bd{G}}^{(4B)}(\bd{3};\bd{4};\bd{2};\bd{1})
:  \beta_{\bd{3}} \alpha^\dagger_{-\bd{4}} \alpha_{\bd{2}}   \beta^\dagger_{-\bd{1}}  : \nonumber\\
&+\frac{1}{(2!)^2} \Gamma_{\bd{G}}^{(5A)}(\bd{3},\bd{4};\bd{2},\bd{1}) \alpha^\dagger_{-\bd{3}} \alpha^\dagger_{-\bd{4}} \beta^\dagger_{-\bd{2}} \beta^\dagger_{-\bd{1}}+\frac{1}{(2!)^2} \Gamma_{\bd{G}}^{(5B)}(\bd{3},\bd{4};\bd{2},\bd{1}) \beta_{\bd{3}} \beta_{\bd{4}} \alpha_{\bd{2}} \alpha_{\bd{1}} \bigg\}.
\label{eq:quarticHamiltonian}
\end{align}
Here the symbol $: \ldots :$ generates normal-ordering of the $\beta$-bosons, for example
 \begin{equation}
 : \beta_{\bd{3}} \beta_{\bd{4}}   \beta^\dagger_{-\bd{2}} \beta^\dagger_{-\bd{1}}  :
 \; =   \beta^\dagger_{-\bd{2}} \beta^\dagger_{-\bd{1}}  \beta_{\bd{3}} \beta_{\bd{4}} .
 \end{equation}
Note that the Hamiltonian in Eq.~(\ref{eq:quarticHamiltonian}) is already normal-ordered in the $\alpha$-bosons. The reason why we prefer to use normal-ordering symbols to write 
down the right-hand side of  Eq.~(\ref{eq:quarticHamiltonian})  is that
then the momentum labels of all vertices and operators appear in the order $\bd{3}, \bd{4} , \bd{2} , \bd{1}$. The fact that the labeling in ${\cal{H}}_4$ is most symmetric if the
$\alpha$-bosons are normal-ordered while the $\beta$-bosons are anti-normal-ordered
has been pointed out long time ago by Harris {\it{el al.}} in Ref.~[\onlinecite{Harris71}].
In Eq.~\eqref{eq:quarticHamiltonian} 
we have explicitly symmetrized the vertices with respect to the permutation of all labels attached to operators of the same type; the combinatorial prefactors
are introduced to cancel combinatorial factors arising in the perturbative treatment of the interaction. The vertices are symmetric with respect to the permutation of all labels separated by a comma. 
Explicitly, the properly symmetrized vertices can be written in the following form. The first two vertices are symmetric with respect to the independent exchange of $\bd{3} \leftrightarrow \bd{4}$ and $\bd{2} \leftrightarrow \bd{1}$,
\begin{align}
        \Gamma_{\bd{G}}^{(1A)}(\bd{3},\bd{4};\bd{2},\bd{1}) {}\equiv{} & \Gamma_{\bd{G}}^{\bar{\alpha}\bar{\alpha}\alpha \alpha}(\bd{3},\bd{4};\bd{2},\bd{1})\nonumber\\
        {}={}&(u_{\bd{3}}v_{\bd{4}}^* W_{\bd{2}+\bd{4}}+v_{\bd{3}}^* u_{\bd{4}} W_{\bd{2}+\bd{3}})v_{\bd{2}} u_{\bd{1}}^* + (u_{\bd{3}}v_{\bd{4}}^* W_{\bd{1}+\bd{4}}+v_{\bd{3}}^* u_{\bd{4}} W_{\bd{1}+\bd{3}}) u_{\bd{2}}^* v_{\bd{1}} \nonumber\\
        & + s_{\bd{G}} (v_{\bd{3}}^* u_{\bd{4}} W_{\bd{2}+\bd{4}}+u_{\bd{3}} v_{\bd{4}}^* W_{\bd{2}+\bd{3}}) u_{\bd{2}}^* v_{\bd{1}} + s_{\bd{G}} (v_{\bd{3}}^* u_{\bd{4}} W_{\bd{1}+\bd{4}}+u_{\bd{3}} v_{\bd{4}}^* W_{\bd{1}+\bd{3}}) v_{\bd{2}} u_{\bd{1}}^* \nonumber\\
        & - u_{\bd{3}} u_{\bd{4}} ( W_{\bd{1}} u_{\bd{2}}^* v_{\bd{1}} + W_{\bd{2}} v_{\bd{2}} u_{\bd{1}}^*) -s_{\bd{G}} v_{\bd{3}}^* v_{\bd{4}}^* (W_{\bd{1}} v_{\bd{2}} u_{\bd{1}}^* + W_{\bd{2}}) u_{\bd{2}}^* v_{\bd{1}}\nonumber\\
        &  - (v_{\bd{3}}^* u_{\bd{4}} W_{\bd{3}}+ u_{\bd{3}} v_{\bd{4}}^* W_{\bd{4}}) u_{\bd{2}}^* u_{\bd{1}}^* -s_{\bd{G}} (u_{\bd{3}} v_{\bd{4}}^* W_{\bd{3}} +  v_{\bd{3}}^* u_{\bd{4}} W_{\bd{4}}) v_{\bd{2}} v_{\bd{1}}\nonumber\\
        &+u_{\bd{3}} u_{\bd{4}} u_{\bd{2}}^* u_{\bd{1}}^* W_{\bd{34;21}}^A + s_{\bd{G}} v_{\bd{3}}^* v_{\bd{4}}^* v_{\bd{2}} v_{\bd{1}} W_{\bd{34;21}}^B,
\end{align}
\begin{align}
        \Gamma_{\bd{G}}^{(1B)}(\bd{3},\bd{4};\bd{2},\bd{1}) {}\equiv{}& \Gamma_{\bd{G}}^{\beta \beta \bar{\beta} \bar{\beta}}(\bd{3},\bd{4};\bd{2},\bd{1})=  \{ \text{exchange}\ u \leftrightarrow v\ \text{in} \ \Gamma^{(1A)} \} \nonumber\\
        {}={}&(v_{\bd{3}}u_{\bd{4}}^* W_{\bd{2}+\bd{4}}+u_{\bd{3}}^* v_{\bd{4}} W_{\bd{2}+\bd{3}})u_{\bd{2}} v_{\bd{1}}^* + (v_{\bd{3}}u_{\bd{4}}^* W_{\bd{1}+\bd{4}}+u_{\bd{3}}^* v_{\bd{4}} W_{\bd{1}+\bd{3}}) v_{\bd{2}}^* u_{\bd{1}} \nonumber\\
        & + s_{\bd{G}} (u_{\bd{3}}^* v_{\bd{4}} W_{\bd{2}+\bd{4}}+v_{\bd{3}} u_{\bd{4}}^* W_{\bd{2}+\bd{3}}) v_{\bd{2}}^* u_{\bd{1}} + s_{\bd{G}} (u_{\bd{3}}^* v_{\bd{4}} W_{\bd{1}+\bd{4}}+v_{\bd{3}} u_{\bd{4}}^* W_{\bd{1}+\bd{3}}) u_{\bd{2}} v_{\bd{1}}^* \nonumber\\
        & - v_{\bd{3}} v_{\bd{4}} ( W_{\bd{1}} v_{\bd{2}}^* u_{\bd{1}} + W_{\bd{2}} u_{\bd{2}} v_{\bd{1}}^*) -s_{\bd{G}} u_{\bd{3}}^* u_{\bd{4}}^* (W_{\bd{1}} u_{\bd{2}} v_{\bd{1}}^* + W_{\bd{2}}) v_{\bd{2}}^* u_{\bd{1}}\nonumber\\
        &  - (u_{\bd{3}}^* v_{\bd{4}} W_{\bd{3}}+ v_{\bd{3}} u_{\bd{4}}^* W_{\bd{4}}) v_{\bd{2}}^* v_{\bd{1}}^* -s_{\bd{G}} (v_{\bd{3}} u_{\bd{4}}^* W_{\bd{3}} +  u_{\bd{3}}^* v_{\bd{4}} W_{\bd{4}}) u_{\bd{2}} u_{\bd{1}}\nonumber\\
        &+v_{\bd{3}} v_{\bd{4}} v_{\bd{2}}^* v_{\bd{1}}^* W_{\bd{34;21}}^A + s_{\bd{G}} u_{\bd{3}}^* u_{\bd{4}}^* u_{\bd{2}} u_{\bd{1}} W_{\bd{34;21}}^B.
\end{align}
The next two vertices are only symmetric with respect to the exchange of the right two labels, $\bd{1} \leftrightarrow \bd{2}$,
\begin{align}
        \Gamma_{\bd{G}}^{(2A)}(\bd{3};\bd{4};\bd{2},\bd{1}){}\equiv{}& \Gamma_{\bd{G}}^{\bar{\alpha}\beta \alpha \alpha}(\bd{3};\bd{4};\bd{2},\bd{1})\nonumber\\
        {}={}& -(u_{\bd{3}}u_{\bd{4}}^* W_{\bd{2}+\bd{4}}+v_{\bd{3}}^* v_{\bd{4}} W_{\bd{2}+\bd{3}}) v_{\bd{2}} u_{\bd{1}}^* -(u_{\bd{3}}u_{\bd{4}}^* W_{\bd{1}+\bd{4}}+v_{\bd{3}}^* v_{\bd{4}} W_{\bd{1}+\bd{3}}) u_{\bd{2}}^* v_{\bd{1}} \nonumber\\
        & - s_{\bd{G}} (v_{\bd{3}}^* v_{\bd{4}} W_{\bd{2}+\bd{4}}+u_{\bd{3}} u_{\bd{4}}^* W_{\bd{2}+\bd{3}}) u_{\bd{2}}^* v_{\bd{1}} - s_{\bd{G}} (v_{\bd{3}}^* v_{\bd{4}} W_{\bd{1}+\bd{4}}+u_{\bd{3}} u_{\bd{4}}^* W_{\bd{1}+\bd{3}}) v_{\bd{2}} u_{\bd{1}}^* \nonumber\\
        & + u_{\bd{3}} v_{\bd{4}} ( W_{\bd{1}} u_{\bd{2}}^* v_{\bd{1}} + W_{\bd{2}} v_{\bd{2}} u_{\bd{1}}^*) + s_{\bd{G}} v_{\bd{3}}^* u_{\bd{4}}^* (W_{\bd{1}} v_{\bd{2}} u_{\bd{1}}^* + W_{\bd{2}} u_{\bd{2}}^* v_{\bd{1}})\nonumber\\
        &  + (v_{\bd{3}}^* v_{\bd{4}}  W_{\bd{3}}+ u_{\bd{3}} u_{\bd{4}}^* W_{\bd{4}}) u_{\bd{2}}^* u_{\bd{1}}^* + s_{\bd{G}} (u_{\bd{3}} u_{\bd{4}}^* W_{\bd{3}} +  v_{\bd{3}}^* v_{\bd{4}} W_{\bd{4}}) v_{\bd{2}} v_{\bd{1}}\nonumber\\
        & -u_{\bd{3}} v_{\bd{4}} u_{\bd{2}}^* u_{\bd{1}}^* W_{\bd{34;21}}^A - s_{\bd{G}} v_{\bd{3}}^* u_{\bd{4}}^* v_{\bd{2}} v_{\bd{1}} W_{\bd{34;21}}^B,
\end{align}
\begin{align}
        \Gamma_{\bd{G}}^{(2B)}(\bd{3},\bd{4};\bd{2},\bd{1}){}\equiv{}& \Gamma_{\bd{G}}^{\beta \bar{\alpha} \bar{\beta} \bar{\beta}}(\bd{3};\bd{4};\bd{2},\bd{1})= \{ \text{exchange}\ u \leftrightarrow v\ \text{in}  \ \Gamma^{(2A)} \}\nonumber\\
        {}={}& -(v_{\bd{3}}v_{\bd{4}}^* W_{\bd{2}+\bd{4}}+u_{\bd{3}}^* u_{\bd{4}} W_{\bd{2}+\bd{3}}) u_{\bd{2}} v_{\bd{1}}^* -(v_{\bd{3}}v_{\bd{4}}^* W_{\bd{1}+\bd{4}}+u_{\bd{3}}^* u_{\bd{4}} W_{\bd{1}+\bd{3}}) v_{\bd{2}}^* u_{\bd{1}} \nonumber\\
        & - s_{\bd{G}} (u_{\bd{3}}^* u_{\bd{4}} W_{\bd{2}+\bd{4}}+v_{\bd{3}} v_{\bd{4}}^* W_{\bd{2}+\bd{3}}) v_{\bd{2}}^* u_{\bd{1}} - s_{\bd{G}} (u_{\bd{3}}^* u_{\bd{4}} W_{\bd{1}+\bd{4}}+v_{\bd{3}} v_{\bd{4}}^* W_{\bd{1}+\bd{3}}) u_{\bd{2}} v_{\bd{1}}^* \nonumber\\
        & + v_{\bd{3}} u_{\bd{4}} ( W_{\bd{1}} v_{\bd{2}}^* u_{\bd{1}} + W_{\bd{2}} u_{\bd{2}} v_{\bd{1}}^*) + s_{\bd{G}} u_{\bd{3}}^* v_{\bd{4}}^* (W_{\bd{1}} u_{\bd{2}} v_{\bd{1}}^* + W_{\bd{2}} v_{\bd{2}}^* u_{\bd{1}})\nonumber\\
        &  + (u_{\bd{3}}^* u_{\bd{4}}  W_{\bd{3}}+ v_{\bd{3}} v_{\bd{4}}^* W_{\bd{4}}) v_{\bd{2}}^* v_{\bd{1}}^* + s_{\bd{G}} (v_{\bd{3}} v_{\bd{4}}^* W_{\bd{3}} +  u_{\bd{3}}^* u_{\bd{4}} W_{\bd{4}}) u_{\bd{2}} u_{\bd{1}}\nonumber\\
        & -v_{\bd{3}} u_{\bd{4}} v_{\bd{2}}^* v_{\bd{1}}^* W_{\bd{34;21}}^A - s_{\bd{G}} u_{\bd{3}}^* v_{\bd{4}}^* u_{\bd{2}} u_{\bd{1}} W_{\bd{34;21}}^B.
\end{align}
The following two vertices are symmetric with respect to the exchange of the left two labels, $\bd{3} \leftrightarrow \bd{4}$,
\begin{align}
        \Gamma_{\bd{G}}^{(3A)}(\bd{3},\bd{4};\bd{2};\bd{1}){}\equiv{}& \Gamma_{\bd{G}}^{\bar{\alpha}\bar{\alpha}\alpha \bar{\beta}}(\bd{3},\bd{4};\bd{2};\bd{1})\nonumber\\
        {}={}& -u_{\bd{3}}v_{\bd{4}}^*( W_{\bd{2}+\bd{4}}v_{\bd{2}} v_{\bd{1}}^*+ W_{\bd{1}+\bd{4}} u_{\bd{2}}^* u_{\bd{1}} ) -v_{\bd{3}}^* u_{\bd{4}}( W_{\bd{2}+\bd{3}} v_{\bd{2}} v_{\bd{1}}^* +W_{\bd{1}+\bd{3}} v_{\bd{2}}^* v_{\bd{1}}) \nonumber\\
        & - s_{\bd{G}} v_{\bd{3}}^* u_{\bd{4}} ( W_{\bd{2}+\bd{4}} u_{\bd{2}}^* u_{\bd{1}} + W_{\bd{1}+\bd{4}} v_{\bd{2}} v_{\bd{1}}^*) -s_{\bd{G}} u_{\bd{3}} v_{\bd{4}}^* ( W_{\bd{2}+\bd{3}} u_{\bd{2}}^* u_{\bd{1}}+ W_{\bd{1}+\bd{3}} u_{\bd{2}}^* u_{\bd{1}}) \nonumber\\
        & + u_{\bd{3}} u_{\bd{4}} ( W_{\bd{1}} u_{\bd{2}}^* u_{\bd{1}} + W_{\bd{2}} v_{\bd{2}}  v_{\bd{1}}^*) + s_{\bd{G}} v_{\bd{3}}^* v_{\bd{4}}^* (W_{\bd{1}} v_{\bd{2}} v_{\bd{1}}^* + W_{\bd{2}} u_{\bd{2}}^* u_{\bd{1}})\nonumber\\
        &  + (v_{\bd{3}}^* u_{\bd{4}}  W_{\bd{3}}+ u_{\bd{3}} v_{\bd{4}}^* W_{\bd{4}}) u_{\bd{2}}^* v_{\bd{1}}^* + s_{\bd{G}} (u_{\bd{3}} v_{\bd{4}}^* W_{\bd{3}} +  v_{\bd{3}}^* u_{\bd{4}} W_{\bd{4}}) v_{\bd{2}} u_{\bd{1}}\nonumber\\
        &-u_{\bd{3}} u_{\bd{4}} u_{\bd{2}}^* v_{\bd{1}}^* W_{\bd{34;21}}^A - s_{\bd{G}} v_{\bd{3}}^* v_{\bd{4}}^* v_{\bd{2}} u_{\bd{1}} W_{\bd{34;21}}^B,
\end{align}
\begin{align}
        \Gamma_{\bd{G}}^{(3B)}(\bd{3},\bd{4};\bd{2};\bd{1}){}\equiv{}& \Gamma_{\bd{G}}^{\beta \beta \bar{\beta} \alpha}(\bd{3},\bd{4};\bd{2};\bd{1})= \{ \text{exchange}\ u \leftrightarrow v\ \text{in}  \ \Gamma^{(3A)} \}\nonumber\\
        {}={}& -v_{\bd{3}}u_{\bd{4}}^*( W_{\bd{2}+\bd{4}}u_{\bd{2}} u_{\bd{1}}^*+ W_{\bd{1}+\bd{4}} v_{\bd{2}}^* v_{\bd{1}}) -u_{\bd{3}}^* v_{\bd{4}}( W_{\bd{2}+\bd{3}} u_{\bd{2}} u_{\bd{1}}^* +W_{\bd{1}+\bd{3}} u_{\bd{2}}^* v_{\bd{1}}) \nonumber\\
        & - s_{\bd{G}} u_{\bd{3}}^* v_{\bd{4}} ( W_{\bd{2}+\bd{4}} v_{\bd{2}}^* v_{\bd{1}} + W_{\bd{1}+\bd{4}} u_{\bd{2}} u_{\bd{1}}^*) -s_{\bd{G}} v_{\bd{3}} u_{\bd{4}}^* ( W_{\bd{2}+\bd{3}} v_{\bd{2}}^* v_{\bd{1}}+ W_{\bd{1}+\bd{3}} v_{\bd{2}}^* v_{\bd{1}} )\nonumber\\
        & + v_{\bd{3}} v_{\bd{4}} ( W_{\bd{1}} v_{\bd{2}}^* v_{\bd{1}} + W_{\bd{2}} u_{\bd{2}}  u_{\bd{1}}^*) + s_{\bd{G}} u_{\bd{3}}^* u_{\bd{4}}^* (W_{\bd{1}} u_{\bd{2}} u_{\bd{1}}^* + W_{\bd{2}} v_{\bd{2}}^* v_{\bd{1}})\nonumber\\
        &  + (u_{\bd{3}}^* v_{\bd{4}} W_{\bd{3}}+ v_{\bd{3}} u_{\bd{4}}^* W_{\bd{4}}) v_{\bd{2}}^* u_{\bd{1}}^* + s_{\bd{G}} (v_{\bd{3}} u_{\bd{4}}^* W_{\bd{3}} +  u_{\bd{3}}^* v_{\bd{4}} W_{\bd{4}}) u_{\bd{2}} v_{\bd{1}}\nonumber\\
        &-v_{\bd{3}} v_{\bd{4}} v_{\bd{2}}^* u_{\bd{1}}^* W_{\bd{34;21}}^A - s_{\bd{G}} u_{\bd{3}}^* u_{\bd{4}}^* u_{\bd{2}} v_{\bd{1}} W_{\bd{34;21}}^B.
\end{align}
Then, there are two vertices without any permutation symmetries of the external labels because all external legs are associated with different operators,
\begin{align}
        \Gamma_{\bd{G}}^{(4A)}(\bd{3};\bd{4};\bd{2},\bd{1}){}\equiv{}& \Gamma_{\bd{G}}^{\bar{\alpha}\beta \bar{\beta} \alpha}(\bd{3};\bd{4};\bd{2},\bd{1})\nonumber\\
        {}={}& 2u_{\bd{3}}u_{\bd{4}}^*( W_{\bd{2}+\bd{4}}u_{\bd{2}} u_{\bd{1}}^* + W_{\bd{1}+\bd{4}} v_{\bd{2}}^* v_{\bd{1}} ) - 2 s_{\bd{G}} v_{\bd{3}}^* v_{\bd{4}} ( W_{\bd{2}+\bd{4}} v_{\bd{2}}^* v_{\bd{1}} + W_{\bd{1}+\bd{4}} u_{\bd{2}} u_{\bd{1}}^*) \nonumber\\
        & - u_{\bd{3}} v_{\bd{4}} ( W_{\bd{1}} v_{\bd{2}}^* v_{\bd{1}} + W_{\bd{2}} u_{\bd{2}}  u_{\bd{1}}^*) + s_{\bd{G}} v_{\bd{3}}^* u_{\bd{4}}^* (W_{\bd{2}} u_{\bd{2}} u_{\bd{1}}^* + W_{\bd{2}} v_{\bd{2}}^* v_{\bd{1}})\nonumber\\
        &  - (v_{\bd{3}}^* v_{\bd{4}}  W_{\bd{3}}+ u_{\bd{3}} u_{\bd{4}}^* W_{\bd{4}}) v_{\bd{2}}^* u_{\bd{1}}^* - s_{\bd{G}} (u_{\bd{3}} u_{\bd{4}}^* W_{\bd{3}} +  v_{\bd{3}}^* v_{\bd{4}} W_{\bd{4}}) u_{\bd{2}} v_{\bd{1}}\nonumber\\
        & + u_{\bd{3}} v_{\bd{4}} v_{\bd{2}}^* u_{\bd{1}}^* W_{\bd{34;21}}^A + s_{\bd{G}} v_{\bd{3}}^* u_{\bd{4}}^* u_{\bd{2}} v_{\bd{1}} W_{\bd{34;21}}^B,
\end{align}
\begin{align}
        \Gamma_{\bd{G}}^{(4B)}(\bd{3};\bd{4};\bd{2},\bd{1}){}\equiv{}& \Gamma_{\bd{G}}^{\beta \bar{\alpha} \alpha \bar{\beta}}(\bd{3};\bd{4};\bd{2},\bd{1})=  \{ \text{exchange}\ u \leftrightarrow v\ \text{in}\  \Gamma^{(4A)} \}\nonumber\\
        {}={}& 2v_{\bd{3}}v_{\bd{4}}^*( W_{\bd{2}+\bd{4}}v_{\bd{2}} v_{\bd{1}}^* + W_{\bd{1}+\bd{4}} u_{\bd{2}}^* u_{\bd{1}} ) - 2 s_{\bd{G}} u_{\bd{3}}^* u_{\bd{4}} ( W_{\bd{2}+\bd{4}} u_{\bd{2}}^* u_{\bd{1}} + W_{\bd{1}+\bd{4}} v_{\bd{2}} v_{\bd{1}}^*) \nonumber\\
        & - v_{\bd{3}} u_{\bd{4}} ( W_{\bd{1}} u_{\bd{2}}^* u_{\bd{1}} + W_{\bd{2}} v_{\bd{2}}  v_{\bd{1}}^*) + s_{\bd{G}} u_{\bd{3}}^* v_{\bd{4}}^* (W_{\bd{2}} v_{\bd{2}} v_{\bd{1}}^* + W_{\bd{2}} u_{\bd{2}}^* u_{\bd{1}})\nonumber\\
        &  - (u_{\bd{3}}^* u_{\bd{4}}  W_{\bd{3}}+ v_{\bd{3}} v_{\bd{4}}^* W_{\bd{4}}) u_{\bd{2}}^* v_{\bd{1}}^* - s_{\bd{G}} (v_{\bd{3}} v_{\bd{4}}^* W_{\bd{3}} +  u_{\bd{3}}^* u_{\bd{4}} W_{\bd{4}}) v_{\bd{2}} u_{\bd{1}}\nonumber\\
        & + v_{\bd{3}} u_{\bd{4}} u_{\bd{2}}^* v_{\bd{1}}^* W_{\bd{34;21}}^A + s_{\bd{G}} u_{\bd{3}}^* v_{\bd{4}}^* v_{\bd{2}} u_{\bd{1}} W_{\bd{34;21}}^B.
\end{align}
Finally, the last two vertices, which are anomalous, are, like the first two vertices, symmetric with respect to the independent exchange of $\bd{3} \leftrightarrow \bd{4}$ and $\bd{2} \leftrightarrow \bd{1}$,
\begin{align}
        \Gamma_{\bd{G}}^{(5A)}(\bd{3};\bd{4};\bd{2},\bd{1}) {}\equiv{}& \Gamma_{\bd{G}}^{\bar{\alpha}\bar{\alpha}\bar{\beta} \bar{\beta}}(\bd{3};\bd{4};\bd{2},\bd{1})\nonumber\\
        {}={}& (u_{\bd{3}}v_{\bd{4}}^* W_{\bd{2}+\bd{4}}+v_{\bd{3}}^* u_{\bd{4}} W_{\bd{2}+\bd{3}})u_{\bd{2}} v_{\bd{1}}^* + (u_{\bd{3}}v_{\bd{4}}^* W_{\bd{1}+\bd{4}}+v_{\bd{3}}^* u_{\bd{4}} W_{\bd{1}+\bd{3}}) v_{\bd{2}}^* u_{\bd{1}} \nonumber\\
        & + s_{\bd{G}} (v_{\bd{3}}^* u_{\bd{4}} W_{\bd{2}+\bd{4}}+u_{\bd{3}} v_{\bd{4}}^* W_{\bd{2}+\bd{3}}) v_{\bd{2}}^* u_{\bd{1}} + s_{\bd{G}} (v_{\bd{3}}^* u_{\bd{4}} W_{\bd{1}+\bd{4}}+u_{\bd{3}} v_{\bd{4}}^* W_{\bd{1}+\bd{3}}) u_{\bd{2}} v_{\bd{1}}^* \nonumber\\
        & - u_{\bd{3}} u_{\bd{4}} ( W_{\bd{1}} v_{\bd{2}}^* u_{\bd{1}} + W_{\bd{2}} u_{\bd{2}} v_{\bd{1}}^*) -s_{\bd{G}} v_{\bd{3}}^* v_{\bd{4}}^* (W_{\bd{1}} u_{\bd{2}} v_{\bd{1}}^* + W_{\bd{2}}) v_{\bd{2}}^* u_{\bd{1}}\nonumber\\
        &  - (v_{\bd{3}}^* u_{\bd{4}} W_{\bd{3}}+ u_{\bd{3}} v_{\bd{4}}^* W_{\bd{4}}) v_{\bd{2}}^* v_{\bd{1}}^* -s_{\bd{G}} (u_{\bd{3}} v_{\bd{4}}^* W_{\bd{3}} +  v_{\bd{3}}^* u_{\bd{4}} W_{\bd{4}}) u_{\bd{2}} u_{\bd{1}}\nonumber\\
        &+u_{\bd{3}} u_{\bd{4}} v_{\bd{2}}^* v_{\bd{1}}^* W_{\bd{34;21}}^A + s_{\bd{G}} v_{\bd{3}}^* v_{\bd{4}}^* u_{\bd{2}} u_{\bd{1}} W_{\bd{34;21}}^B,
\end{align}
\begin{align}
        \Gamma_{\bd{G}}^{(5B)}(\bd{3};\bd{4};\bd{2},\bd{1}){} \equiv {}& \Gamma_{\bd{G}}^{\beta \beta \alpha \alpha}(\bd{3};\bd{4};\bd{2},\bd{1}) =  \{ \text{exchange}\ u \leftrightarrow v\ \text{in}\  \Gamma^{(5A)} \}  \nonumber\\
        {}={}& (v_{\bd{3}}u_{\bd{4}}^* W_{\bd{2}+\bd{4}}+u_{\bd{3}}^* v_{\bd{4}} W_{\bd{2}+\bd{3}})v_{\bd{2}} u_{\bd{1}}^* + (v_{\bd{3}}u_{\bd{4}}^* W_{\bd{1}+\bd{4}}+u_{\bd{3}}^* v_{\bd{4}} W_{\bd{1}+\bd{3}}) u_{\bd{2}}^* v_{\bd{1}} \nonumber\\
        & + s_{\bd{G}} (u_{\bd{3}}^* v_{\bd{4}} W_{\bd{2}+\bd{4}}+ v_{\bd{3}} u_{\bd{4}}^* W_{\bd{2}+\bd{3}}) u_{\bd{2}}^* v_{\bd{1}} + s_{\bd{G}} (u_{\bd{3}}^* v_{\bd{4}} W_{\bd{1}+\bd{4}} + v_{\bd{3}} u_{\bd{4}}^* W_{\bd{1}+\bd{3}}) v_{\bd{2}} u_{\bd{1}}^* \nonumber\\
        & - v_{\bd{3}} v_{\bd{4}} ( W_{\bd{1}} u_{\bd{2}}^* v_{\bd{1}} + W_{\bd{2}} v_{\bd{2}} u_{\bd{1}}^*) -s_{\bd{G}} u_{\bd{3}}^* u_{\bd{4}}^* (W_{\bd{1}} v_{\bd{2}} u_{\bd{1}}^* + W_{\bd{2}}) u_{\bd{2}}^* v_{\bd{1}}\nonumber\\
        &  - (u_{\bd{3}}^* v_{\bd{4}} W_{\bd{3}}+ v_{\bd{3}} u_{\bd{4}}^* W_{\bd{4}}) u_{\bd{2}}^* u_{\bd{1}}^* -s_{\bd{G}} (v_{\bd{3}} u_{\bd{4}}^* W_{\bd{3}} +  u_{\bd{3}}^* v_{\bd{4}} W_{\bd{4}}) v_{\bd{2}} v_{\bd{1}}\nonumber\\
        &+v_{\bd{3}} v_{\bd{4}} u_{\bd{2}}^* u_{\bd{1}}^* W_{\bd{34;21}}^A + s_{\bd{G}} u_{\bd{3}}^* u_{\bd{4}}^* v_{\bd{2}} v_{\bd{1}} W_{\bd{34;21}}^B.
\end{align}
\end{widetext}

\section*{APPENDIX B: Spin-wave corrections at zero temperature}
\setcounter{equation}{0}
\renewcommand{\theequation}{B\arabic{equation}}

In this appendix, we give some  technical details
of the spin-wave calculations presented in Sec.~\ref{sec:spinwavetheory}.
First of all, the
following relations are useful to manipulate some of the terms generated via the $1/S$-expansion in momentum space:
{\allowdisplaybreaks
\begin{subequations}
    \begin{align}
    u_{\bd{p}}^2 v_{\bd{k}}^2+u_{\bd{k}}^2 v_{\bd{p}}^2 &= \frac{(1-\eta_{\bd{p}})(1-\eta_{\bd{k}})-\epsilon_{\bd{k}} \epsilon_{\bd{p}}}{2 \epsilon_{\bd{k}}\epsilon_{\bd{p}}},\\
    u_{\bd{k}} v_{\bd{k}} &= \frac{\gamma_{\bd{k}}}{2 \epsilon_{\bd{k}}},\\
    u_{\bd{k}}^2 + v_{\bd{k}}^2 &= \frac{1-\eta_{\bd{k}}}{\epsilon_{\bd{k
    }}},\\
    u_{\bd{k}}^2 &= \frac{1-\eta_{\bd{k}}+\epsilon_{\bd{k}}}{2\epsilon_{\bd{k}}},\\
    v_{\bd{k}}^2 &= \frac{1-\eta_{\bd{k}}-\epsilon_{\bd{k}}}{2\epsilon_{\bd{k}}}.
\end{align}
\end{subequations}
In addition, we use the relation
\begin{align}
    \sum\limits_{\bd{k}}& \gamma_{\bd{p}-\bd{k}} \gamma_{\bd{k}}= \gamma_{\bd{p}} \sum\limits_{\bd{k}} \gamma_{\bd{k}}^2,
\end{align}
along with the identity
\begin{align}
    \cos (k_x a\pm k_y a){}={}&\cos(k_x a)\cos(k_y a)
    \mp \sin(k_x a)\sin(k_y a).
\end{align}
Since the sine functions are antisymmetric under inversion their contribution vanishes when we multiply them by a symmetric function and then sum over the entire 
Brillouin zone. For later convenience we also define
 \begin{equation}
\lambda_{\bd{k}} =\cos (k_x a) \cos (k_y a).
\end{equation}

}
\subsection{Corrections to the ground state energy}

The leading correction of order $S$ arises from
normal-ordering ${\cal{H}}_2$ defined in 
Eq.~\eqref{eq:H2} after expressing the Holstein-Primakoff bosons in terms of Bogoliubov bosons using  Eqs.~\eqref{eq:Bogoliubov} and \eqref{eq:uv}. A simple calculation gives
\begin{align}
    \frac{E_0^{(1)}}{N}  &=\frac{1}{N}\sum\limits_{\bd{k}} \left\lbrace {\omega_{\bd{k}}^{-}} - 4 J S \left[ 1-\frac{J_2}{J}(1-\lambda_{\bd{k}}^B)\right] \right\rbrace\nonumber\\
    &= -2 J S \frac{2}{N} \sum\limits_{\bd{k}} (1-\eta_{\bd{k}} -\epsilon_{\bd{k}}),
\end{align}
which is Eq.~\eqref{eq:e1} of the main text.

The next-to-leading correction of order $S^0 = 1$ can be obtained from 
${\cal{H}}_4$ defined in Eq.~\eqref{eq:h4holstein} after Bogoliubov transformation and subsequent normal ordering. 
Before applying the Bogoliubov transformation to $\mathcal{H}_4$, it is convenient to normal order the  $a$-bosons and anti-normal order the  $b$-bosons, resulting in
\begin{widetext}
   \begin{align}
\mathcal{H}_4 & {}={} \frac{2}{N} \sum\limits_{\bd{1}\bd{2}\bd{3}\bd{4}\bd{G}} \delta_{\bd{1}+\bd{2}+\bd{3}+\bd{4},\bd{G}} \bigg\{ W_{\bd{2}+\bd{4}} \Big[a_{-\bd{3}}^\dagger b_{\bd{4}} b^\dagger_{\bd{2}} a_{\bd{1}} + s_{\bd{G}} b_{\bd{3}} a^\dagger_{-\bd{4}} a_{\bd{2}} b^\dagger_{\bd{-1}} \Big]
+\frac{1}{2!} W_{\bd{1}} \Big[ a_{-\bd{3}}^\dagger a^\dagger_{\bd{-4}} a_{\bd{2}} b^\dagger_{-\bd{1}} + s_{\bd{G}} b_{\bd{3}}   b_{\bd{4}} b^\dagger_{-\bd{2}} a_{\bd{1}} \Big] \nonumber\\
&\phantom{+\frac{1}{2!} W_{\bd{4}} \Big[} +\frac{1}{2!} W_{\bd{4}} \Big[ a^\dagger_{-\bd{3}}b_{\bd{4}}a_{\bd{2}} a_{\bd{1}} + s_{\bd{G}} b_{\bd{3}} a^\dagger_{-\bd{4}} b^\dagger_{-\bd{2}} b\dagger_{-\bd{1}}  \Big] + \frac{1}{(2!)^2} \left[ W^A_{\bd{3} \bd{4} ; \bd{2} \bd{1}} a_{-\bd{3}}^\dagger a^\dagger_{-\bd{4}} a_{\bd{2}} a_{\bd{1}} + s_{\bd{G}} W^B_{\bd{3} \bd{4} ; \bd{2} \bd{1}} 
b_{-\bd{3}}^\dagger b_{-\bd{4}}^\dagger b_{\bd{2}}^\dagger a_{\bd{1}} \right] \bigg\}\nonumber\\
&\phantom{+\frac{1}{2!} W_{\bd{4}} \Big[}
- \sum_{\bd{k}} \left[ 2 W_0 a_{\bd{k}}^\dagger a_{\bd{k}} +  W_{\bd{k}} \left( a^\dagger_{\bd{k}} b^\dagger_{-\bd{k}} +b_{-\bd{k}} a_{\bd{k}}\right) + \left( \frac{2}{N} \sum\limits_{\bd{p}} W^B_{-\bd{k},-\bd{p};\bd{p},\bd{k}} \right) b_{\bd{k}} b^\dagger_{\bd{k}} \right] + \frac{1}{2}\frac{2}{N} \sum\limits_{\bd{k} \bd{p}} W^B_{-\bd{k},-\bd{p};\bd{p},\bd{k}}.
\label{eq:h4}
\end{align} 
\end{widetext}
This contains additive constants as well as terms which are  quadratic in the $a$- and $b$-bosons,  which contribute to the $1/S^2$-correction to the ground state energy, to the $1/S$-correction to the magnon dispersion, and to the off-diagonal corrections to the magnon self-energies.
Due to the ordering of the bosons in $\mathcal{H}_4$, the Bogoliubov transformation now 
generates an expression which is normal-ordered in the  $\alpha$-bosons and
anti-normal-ordered  in the $\beta$-bosons. Therefore, to bring $\mathcal{H}_4$ into normal-ordered form, it suffices to reorder only the $\beta$-bosons, which gives additional contributions to the aforementioned quantities that contain the vertices. The result of the normal-ordered $: \mathcal{H}_4:$ is given in Eq.~\eqref{eq:quarticHamiltonian} of Appendix A, where we also give the properly symmetrized interaction vertices.
Let us now give the terms generated by the normal ordering. First of all, 
from Eq.~\eqref{eq:h4} we obtain the following three contributions to the $1/S^2$-corrections to the ground state energy,
\begin{align}
    &E_0^{(2)}{}={}E_{0,1}^{(2)} +E_{0,2}^{(2)}+E_{0,3}^{(2)} ,
\end{align}
where the first contribution comes from the last term in Eq.~\eqref{eq:h4},
\begin{align}
    E_{0,1}^{(2)} {}={} & \frac{1}{2}\frac{2}{N}\sum\limits_{\bd{k},\bd{p}}W^B_{-\bd{k},-\bd{p},\bd{p},\bd{k}} = J_2 N \left[\frac{2}{N}\sum\limits_{\bd{k}} (1-\lambda_{\bd{k}})  \right]^2,
    \end{align}
    the second contribution comes from the quadratic terms in the $a$- and $b$-bosons,
    \begin{align}
     E_{0,2}^{(2)} = {} & \sum\limits_{\bd{k}} \bigg[\! -2 W_0 v_{\bd{k}}^2 \!+\! 2 W_{\bd{k}} u_{\bd{k}}v_{\bd{k}} \! - \! \frac{2}{N}\! \sum\limits_{\bd{p}}  W^B_{-\bd{k},-\bd{p},\bd{p}, \bd{k}} u_{\bd{k}}^2  \bigg]\nonumber\\
= {} &2 J \sum\limits_{\bd{k}} \left( -1 +\frac{1-\eta_{\bd{k}}-\gamma_{\bd{k}}^2}{\epsilon_{\bd{k}}} \right)\nonumber\\
& -2 J_2 \frac{2}{N} \sum\limits_{\bd{k}, \bd{p}} \frac{1-\eta_{\bd{k}}+\epsilon_{\bd{k}}}{\epsilon_{\bd{k}}}(1-\lambda_{\bd{k}})(1-\lambda_{\bd{p}}),
    \end{align}
and the last contribution comes from reordering the $\beta$-bosons,
\begin{align}
    E_{0,3}^{(2)} ={}&\frac{1}{2}\frac{2}{N}\sum\limits_{\bd{k},\bd{p}}\Gamma^{(1B)}_0(\bd{k},\bd{p},-\bd{p},-\bd{k}) \nonumber\\
      ={}& \frac{1}{2}\frac{2}{N} \sum\limits_{\bd{k}, \bd{p}} \bigg\{ 2J \Big[1-\frac{(1-\eta_{\bd{k}}-\gamma_{\bd{k}}^2) (1-\eta_{\bd{p}}-\gamma_{\bd{p}}^2)}{\epsilon_{\bd{k}} \epsilon_{\bd{p}} } \Big]\nonumber\\
    &+ 2J_2\frac{(1-\eta_{\bd{k}})(1-\eta_{\bd{p}})+\epsilon_{\bd{k}} \epsilon_{\bd{p}}}{\epsilon_{\bd{k}} \epsilon_{\bd{p}}}  (1-\lambda_{\bd{k}} )(1-\lambda_{\bd{p}}) \bigg\}.
\end{align}
Summing all three contributions yields $E_0^{(2)}$ as given in 
Eq.~\eqref{eq:energycorrection2}.

\subsection{Self-energy corrections to the magnon dispersions}

Consider first the magnon energy with chirality ${p=+}$, which is corrected by the following  two contributions from Eq.~\eqref{eq:h4},
\begin{align}
     {\Sigma_{\bd{k}}^{+}} = {\Sigma_{\bd{k},1}^{+}}+{\Sigma_{\bd{k},2}^{+}}.
\end{align}
Here, the first contribution comes from the quadratic terms in the $a$- and $b$-bosons,
\begin{align}
    {\Sigma_{\bd{k},1}^{+}} ={}&- \bigg[ 2  W_0  u_{\bd{k}}^2 -2 W_{\bd{k}} u_{\bd{k}} v_{\bd{k}} + \frac{2}{N} \sum\limits_{\bd{p}} W^B_{-\bd{k},-\bd{p},\bd{p}, \bd{k}} v_{\bd{k}}^2 \bigg] \nonumber\\
    ={}&  2J \left( 1 + \frac{1-\eta_{\bd{k}}-\gamma_{\bd{k}}^2}{\epsilon_{\bd{k}}} \right)\nonumber\\
    & - \frac{2}{N} \sum\limits_{\bd{p}} \left[\frac{1-\eta_{\bd{k}}-\epsilon_{\bd{k}}}{\epsilon_{\bd{k}}} \right] \Big[2J_2(1-\lambda_{\bd{k}})(1-\lambda_{\bd{p}})\nonumber\\
    &-2J_2^\prime\sin (k_x a) \sin ( k_y a) (1-\lambda_{\bd{p}})\Big],
    \end{align}
    and the second contribution comes from the reordering of the $\beta$-bosons,
    \begin{align}
    {\Sigma_{\bd{k},2}^{+}} ={}&    \frac{1}{N}\! \sum\limits_{\bd{p}}\! \left[ \Gamma^{(4A)}_0 (-\bd{k},-\bd{p},\bd{p},\bd{k})\! +\! \Gamma^{(4B)}_0 (-\bd{p},-\bd{k},\bd{k},\bd{p}) \right]\nonumber\\
={}&-2J \frac{2}{N}\sum\limits_{\bd{p}} \Big\{1+\frac{1}{\epsilon_{\bd{k}}\epsilon_{\bd{p}}}\Big[ (1-\eta_{\bd{k}})(1-\eta_{\bd{p}})\nonumber\\
&+\gamma_{\bd{k}}^2 \gamma_{\bd{p}}^2-\gamma_{\bd{k}}^2(1-\eta_{\bd{p}}) -\gamma_{\bd{p}}^2(1-\eta_{\bd{k}})\Big\}\nonumber\\[0.2cm]
&+ \frac{2}{N} \sum\limits_{\bd{p}} \frac{(1-\eta_{\bd{k}})(1-\eta_{\bd{p}})-\epsilon_{\bd{k}}\epsilon_{\bd{p}}}{ \epsilon_{\bd{k}} \epsilon_{\bd{p}}} \nonumber\\
& \phantom{{}={} \hspace{1cm}} \times 2J_2 (1-\lambda_{\bd{k}})(1-\lambda_{\bd{p}})\nonumber\\[0.2cm]
& + \frac{2}{N}\sum\limits_{\bd{p}} \frac{(1-\eta_{\bd{p}})\epsilon_{\bd{k}}-(1-\eta_{\bd{k}})\epsilon_{\bd{p}}}{ \epsilon_{\bd{k}} \epsilon_{\bd{p}}} \nonumber\\
& \phantom{{}={} \hspace{1cm}} \times  2J_2^\prime\sin (k_x a) \sin(k_y a) (1-\lambda_{\bd{p}}).
\end{align}
The sum of both contributions yields ${{\Sigma}_{\bd{k}}^+}$, given in Eq.~\eqref{eq:dispcorr}.

Next, consider the energy of the magnon with chirality $p=-$, whose energy is renormalized by the following two contributions from Eq.~\eqref{eq:h4},
\begin{align}
     {\Sigma_{\bd{k}}^{-}} = {\Sigma_{\bd{k},1}^{-}}+{\Sigma_{\bd{k},2}^{-}}.
\end{align}
Here, the first contribution comes from the quadratic terms in the $a$- and $b$-bosons,
\begin{align}
    {\Sigma_{\bd{k},1}^{-}} ={}&- 2 W_0  v_{\bd{k}}^2 +2 W_{\bd{k}} u_{\bd{k}} v_{\bd{k}} - \frac{2}{N} \sum\limits_{\bd{p}} W^B_{-\bd{k},-\bd{p},\bd{p}, \bd{k}} u_{\bd{k}}^2 \nonumber\\
     ={} & 2J \left(-1 + \frac{1-\eta_{\bd{k}}-\gamma_{\bd{k}}^2}{\epsilon_{\bd{k}}} \right)\nonumber\\
     &- \frac{2}{N}\sum\limits_{\bd{p}} \frac{1-\eta_{\bd{k}} + \epsilon_{\bd{k}}}{ \epsilon_{\bd{k}}} \Big[ 2J_2(1-\lambda_{\bd{k}})(1-\lambda_{\bd{p}}) \nonumber\\
     & \phantom{{}===={}}- 2J_2^\prime\sin (k_x a) \sin (k_y a) (1-\lambda_{\bd{p}}) \Big],
     \end{align}
     and the second contribution comes from the reordering of the $\beta$-bosons,
     \begin{align}
     {\Sigma_{\bd{k},2}^{-}} ={}&     \frac{2}{N}\sum\limits_{\bd{p}}\Gamma^{(1B)}_0 (\bd{k},-\bd{p},\bd{p},-\bd{k}) \nonumber\\
    ={}&-\frac{4J}{N} \sum\limits_{\bd{p}} \left[ \frac{(1-\eta_{\bd{k}}-\gamma_{\bd{k}}^2) (1-\eta_{\bd{p}}-\gamma_{\bd{p}}^2) }{\epsilon_{\bd{k}} \epsilon_{\bd{p}}} -1\right] \nonumber\\
    & + \frac{2}{N} \sum\limits_{\bd{p}} \frac{(1-\eta_{\bd{k}}-\epsilon_{\bd{k}})(1-\eta_{\bd{p}}-\epsilon_{\bd{p}})}{ 2 \epsilon_{\bd{p}} \epsilon_{\bd{k}}}\nonumber\\
    & \quad \phantom{{}\hspace{-0.05cm}{}}\ \times\Big[ 2J_2 (1-\lambda_{\bd{k}})(1-\lambda_{\bd{p}})\nonumber\\ & \quad \phantom{{}={}} \  \,  + 2J_2^\prime \sin(k_x a) \sin(k_y a) (1-\lambda_{\bd{p}})\Big]\nonumber\\
    & + \frac{2}{N} \sum\limits_{\bd{p}} \frac{(1-\eta_{\bd{k}}+\epsilon_{\bd{k}})(1-\eta_{\bd{p}}+\epsilon_{\bd{p}})}{ 2 \epsilon_{\bd{p}} \epsilon_{\bd{k}}}\nonumber\\
    & \quad \ \phantom{{}\hspace{-0.05cm}{}} \times\Big[ 2J_2 (1-\lambda_{\bd{k}})(1-\lambda_{\bd{p}})\nonumber\\
    &\quad \quad \ \ \,  - 2J_2^\prime\sin(k_x a) \sin(k_y a) (1-\lambda_{\bd{p}})\Big].
\end{align}
The sum of both contributions yields ${{\Sigma}_{\bd{k}}^-}$, given in Eq.~\eqref{eq:dispcorr}.

\subsection{Off-diagonal self-energy}
Finally, let us calculate the off-diagonal self-energy $\Delta_{\bd{k}}$ 
defined via Eq.~\eqref{eq:quadrHamiltonian}.
From Eq.~\eqref{eq:h4} we obtain the following two contributions,
\begin{align}
    \Delta_{\bd{k}} ={}& \Delta_{\bd{k},1}+\Delta_{\bd{k},2},
\end{align}
where the first contribution comes from the quadratic terms in the $a$- and $b$-bosons,
\begin{align}
     \Delta_{\bd{k},1} ={}& 2 W_0 u_{\bd{k}} v_{\bd{k}}\! - \! W_{\bd{k}} \left( u_{\bd{k}}^2 + v_{\bd{k}}^2 \right) 
     \frac{2}{N}\! \sum\limits_{\bd{p}} W^B_{-\bd{k},-\bd{p},\bd{p}, \bd{k}} u_{\bd{k}} v_{\bd{k}}\nonumber\\
        ={} & -2 J \frac{\gamma_{\bd{k}} \eta_{\bd{k}}}{\epsilon_{\bd{k}}} + \frac{\gamma_{\bd{k}}}{ \epsilon_{\bd{k}}} \frac{2}{N} \sum\limits_{\bd{p}} \Big[ 2 J_2 (1-\lambda_{\bd{k}})(1-\lambda_{\bd{p}}) \nonumber\\
        &\phantom{ } - 2 J_2^\prime \sin(k_x) \sin(k_y) (1-\lambda_{\bd{p}}) \Big],
        \end{align}
        and the second contribution comes from the normal-ordering of the $\beta$-bosons,
        \begin{align}
           \Delta_{\bd{k},2} ={}&\frac{2}{N}\!\sum\limits_{\bd{p}} \Gamma^{(2B)}_0(-\bd{p},-\bd{k},\bd{p},\bd{k})\nonumber\\
           ={}& \frac{2}{N}\! \sum\limits_{\bd{p}} \Gamma^{(3B)}_0 (-\bd{p},-\bd{k},\bd{p},\bd{k})\nonumber\\
={} & -\frac{2}{N} \sum\limits_{\bd{p}}\bigg\{-\frac{2J}{\epsilon_{\bd{k}} \epsilon_{\bd{p}}}\bigg[ \gamma_{\bd{k}} \eta_{\bd{k}}(1-\eta_{\bd{p}}-\gamma_{\bd{p}}^2)\bigg]\nonumber\\
     &\quad  \phantom{{}={}} + \left(\frac{\gamma_{\bd{k}}}{\epsilon_{\bd{k}} \epsilon_{\bd{p}} }\right) \bigg[ (1-\eta_{\bd{p}})2 J_2(1-\lambda_{\bd{k}})(1-\lambda_{\bd{p}}) \nonumber\\
     & \phantom{{}={}  \hspace{0.8cm}}- \epsilon_{\bd{p}} 2 J_2^\prime \sin(k_x) \sin(k_y) (1-\lambda_{\bd{p}}) \bigg]\Bigg\},
\end{align}
where the $\Gamma^{(2B)}$-vertex appears in front of the $\alpha^\dagger_{\bd{k}} \beta^\dagger_{-\bd{k}}$ term and the $\Gamma^{(3B)}$-vertex appears in front of the $\alpha_{\bd{k}} \beta_{-\bd{k}}$ term. The result for the off-diagonal self-energy is the sum of both contributions:
\begin{align}
    \Delta_{\bd{k}}{}={}&-2 J \frac{\gamma_{\bd{k}} \eta_{\bd{k}}}{\epsilon_{\bd{k}}} \left[ 1-\frac{2}{N} \sum\limits_{\bd{p}} \frac{1-\eta_{\bd{p}}-\gamma_{\bd{p}}^2}{\epsilon_{\bd{p}}}\right]\nonumber\\
    &- 2 J_2  \frac{\gamma_{\bd{k}}}{\epsilon_{\bd{k}}} (1-\lambda_{\bd{k}})  
    \frac{2}{N}  \sum\limits_{\bd{p}}  \frac{1-\eta_{\bd{p}} -\epsilon_{\bd{p}}}{\epsilon_{\bd{p}}} (1-\lambda_{\bd{p}}),
\end{align}
which is equivalent to Eq.~\eqref{eq:xi} of the main text.
\end{appendix}

\end{document}